%% file: main.tex
\documentclass[journal]{IEEEtran}

\usepackage{graphicx}
\usepackage{latexsym}
\usepackage{CJKutf8}
\usepackage{multirow}
\usepackage{booktabs}
\usepackage{enumitem}
\usepackage{amsmath}
\usepackage{amssymb}
\usepackage{multirow}
\usepackage{array}
\usepackage{xcolor}
\usepackage[boxruled,noend]{algorithm2e}
\usepackage[hidelinks]{hyperref}
\usepackage{tikz}
\usepackage{caption}
\usepackage{subcaption}
\usetikzlibrary{plotmarks}

\newcommand{\inputrating}{\mathbf{X}}
\newcommand{\outputrating}{\hat{\mathbf{X}}}
\newcommand{\predict}{\tilde{\mathbf{X}}}
\newcommand{\sample}{\mathbf{X}^{(in)}}
\newcommand{\real}{\mathcal{R}}
\newcommand{\users}{\mathcal{U}}
\newcommand{\items}{\mathcal{I}}
\newcommand{\select}{\mathcal{S}}
\newcommand{\fakeusers}{\mathcal{V}}
\newcommand{\Lparam}{\Omega}
\newcommand{\Gparam}{\Theta}
\newcommand{\Dparam}{\Phi}
\newcommand{\Sparam}{\Gamma}
\newcommand{\gloss}{\mathcal{L}_{\mathit{gen}}}
\newcommand{\dloss}{\mathcal{L}_{\mathit{dis}}}
\newcommand{\rsloss}{\mathcal{L}_{\mathit{RS}}}
\newcommand{\Gen}{G_{\Gparam}(\sample)}
\newcommand{\Learner}{\mathit{PL}_{\Lparam}(\sample)}

\newcommand{\ours}{Leg-UP\xspace}

\graphicspath{{fig/}}

\begin{document}

\bstctlcite{IEEEexample:BSTcontrol}

\title{Shilling Black-box Recommender Systems by Learning to Generate Fake User Profiles}

\author{Chen~Lin,
        Si~Chen,
        Meifang~Zeng,
        Sheng~Zhang,
        Min~Gao,
        and~Hui~Li
\thanks{Chen Lin is with School of Informatics, Xiamen University, Xiamen 361005, China. E-mail: chenlin@xmu.edu.cn.}
\thanks{Si Chen is with School of Informatics, Xiamen University, Xiamen 361005, China. E-mail: sichen@stu.xmu.edu.cn.}
\thanks{Meifang Zeng is with School of Informatics, Xiamen University, Xiamen 361005, China. E-mail: zengmeifang@stu.xmu.edu.cn.}
\thanks{Sheng Zhang is with School of Informatics, Xiamen University, Xiamen 361005, China. E-mail: zshingjason@163.com.}
\thanks{Min Gao is with School of Big Data and Software Engineering, Chongqing University, Chongqing 400044, China. E-mail: gaomin@cqu.edu.cn.}
\thanks{Hui Li is with School of Informatics, Xiamen University, Xiamen 361005, China. E-mail: hui@xmu.edu.cn. He is the corresponding author.}
}

\maketitle

\begin{abstract}
Due to the pivotal role of Recommender Systems (RS) in guiding customers towards the purchase, there is a natural motivation for unscrupulous parties to spoof RS for profits. In this paper, we study Shilling Attack where an adversarial party injects a number of fake user profiles for improper purposes. Conventional Shilling Attack approaches lack attack transferability (i.e., attacks are not effective on some victim RS models) and/or attack invisibility (i.e., injected profiles can be easily detected). To overcome these issues, we present Leg-UP, a novel attack model based on the Generative Adversarial Network. Leg-UP learns user behavior patterns from real users in the sampled ``templates'' and constructs fake user profiles. To simulate real users, the generator in Leg-UP directly outputs discrete ratings. To enhance attack transferability, the parameters of the generator are optimized by maximizing the attack performance on a surrogate RS model. To improve attack invisibility, Leg-UP adopts a discriminator to guide the generator to generate undetectable fake user profiles. Experiments on benchmarks have shown that Leg-UP exceeds state-of-the-art Shilling Attack methods on a wide range of victim RS models. The source code of our work is available at: https://github.com/XMUDM/ShillingAttack.
\end{abstract}


\IEEEpeerreviewmaketitle

\input{1_intro}
\input{2_related_work}

\input{3_model}

\input{4_imp}
\input{5_learning}

\input{6_experiment}
\input{7_conclude}

\section*{Acknowledgments}
\addcontentsline{toc}{section}{Acknowledgments}
Chen Lin is supported by the National Natural Science Foundation of China (No. 61972328), Alibaba Group through Alibaba Innovative Research program, and Joint Innovation Research Program of Fujian Province of China (No. 2020R0130). Hui Li is supported by the National Natural Science Foundation of China (No. 62002303), the Natural Science Foundation of Fujian Province of China (No. 2020J05001) and the China Fundamental Research Funds for the Central Universities (No. 20720210098).

\bibliographystyle{IEEEtran}
\bibliography{IEEEabrv,ref}

\vspace{-30pt}
\begin{IEEEbiography}[{\vspace{-15pt}\includegraphics[width=1in,height=1.25in,clip,keepaspectratio]{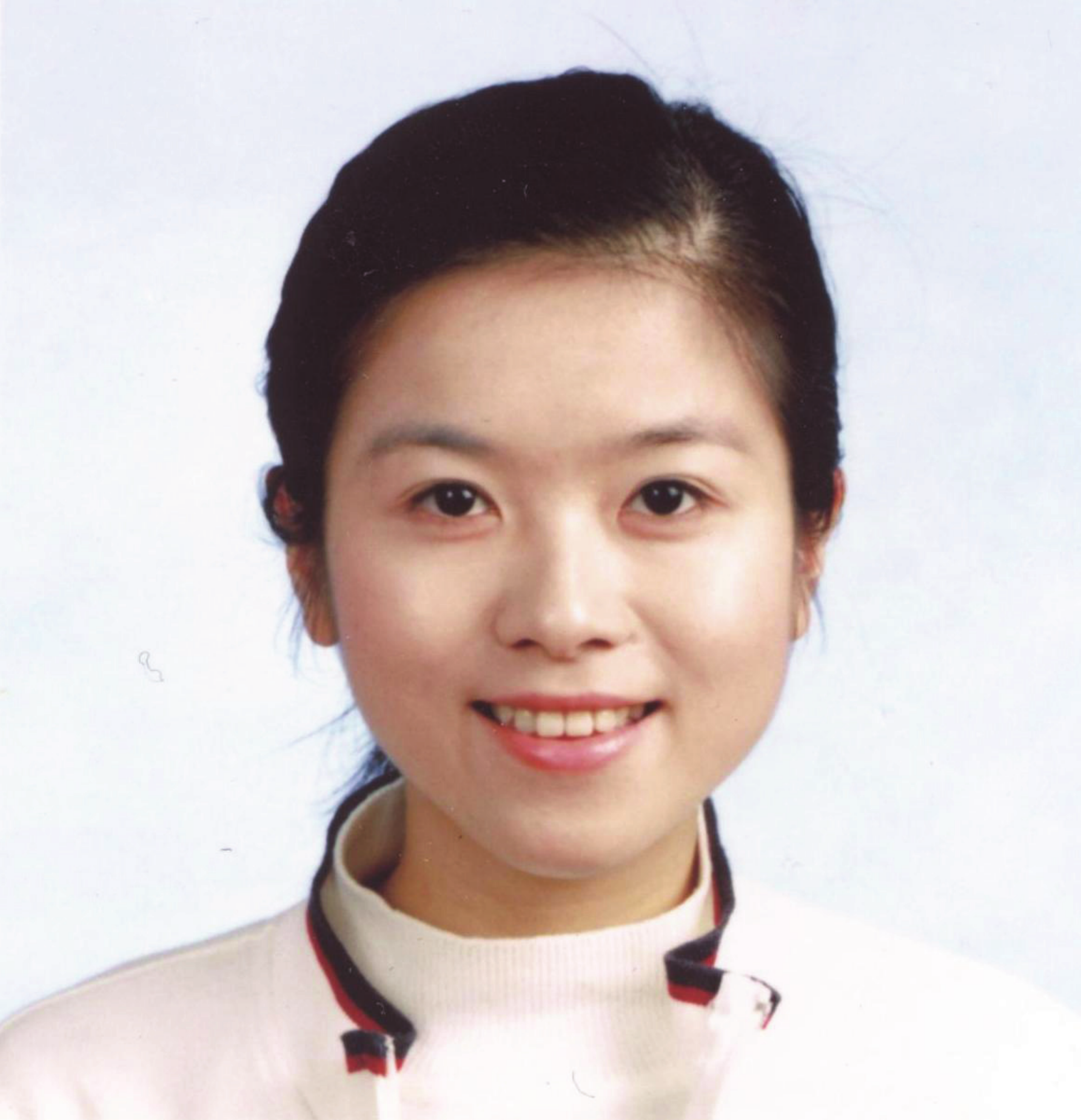}}]
{Chen Lin} received a B. Eng. degree and a Ph.D. degree both from Fudan University, China in 2004 and 2010. She is currently a professor at School of Informatics, Xiamen University, China. Her research interests include web mining and recommender systems.
\end{IEEEbiography}

\vspace{-30pt}
\begin{IEEEbiography}[{\vspace{-20pt}\includegraphics[width=1in,height=1.25in,clip,keepaspectratio]{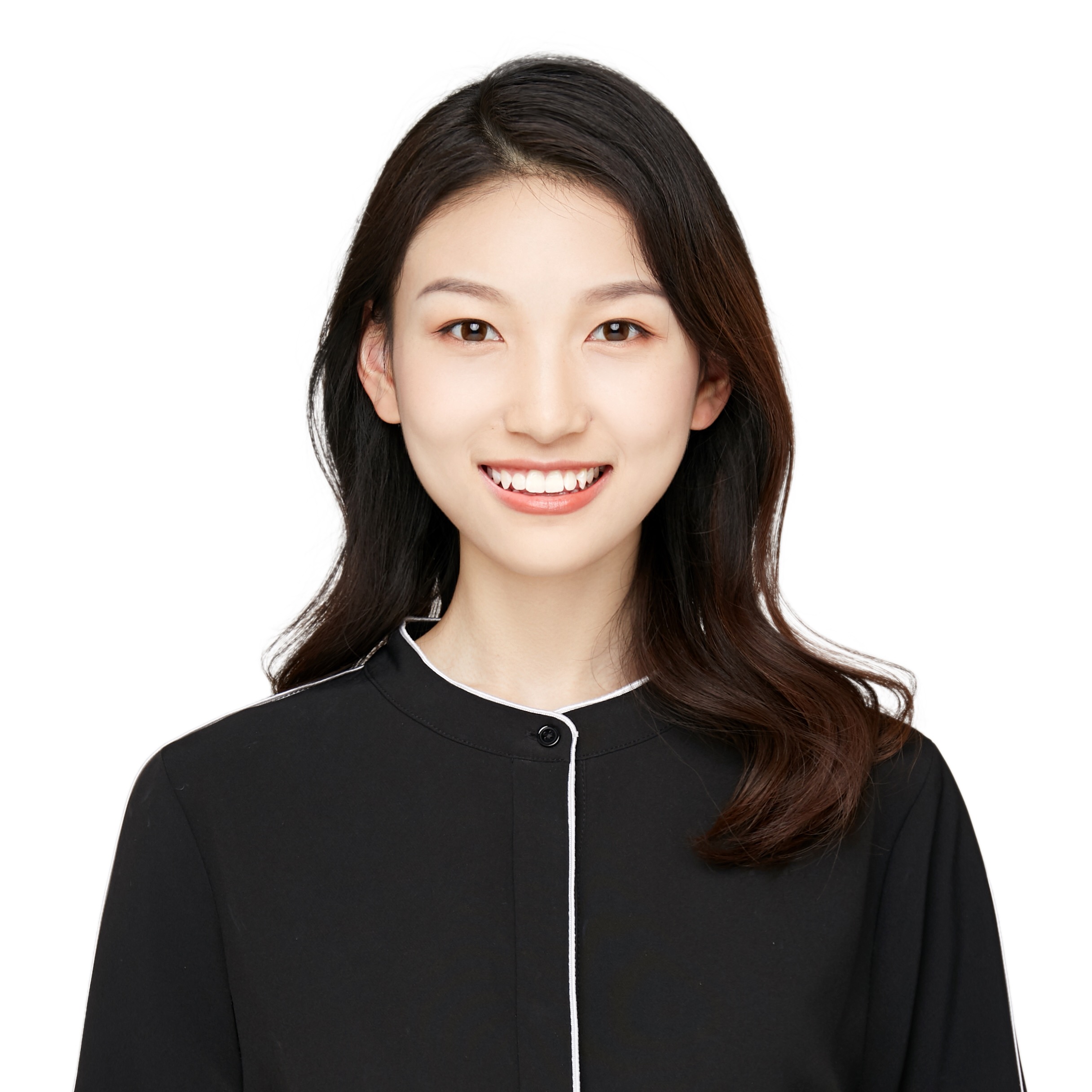}}]
{Si Chen} obtained her master degree from the School of Informatics at Xiamen University. Her research interests include data mining, recommender systems and computational advertising. 
\end{IEEEbiography}

\vspace{-30pt}
\begin{IEEEbiography}[{\vspace{-20pt}\includegraphics[width=1in,height=1.25in,clip,keepaspectratio]{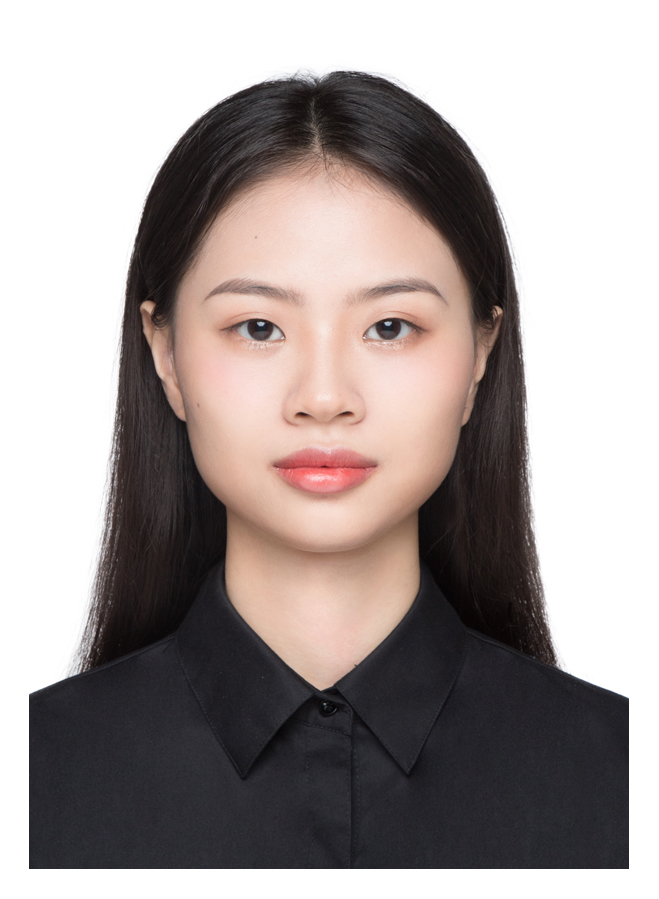}}]
{Meifang Zeng} is a second-year master student in the School of Informatics at Xiamen University. Her research interests include data mining and recommender systems.
\end{IEEEbiography}

\vspace{-30pt}
\begin{IEEEbiography}[{\vspace{-15pt}\includegraphics[width=0.9in,height=1.15in,clip,keepaspectratio]{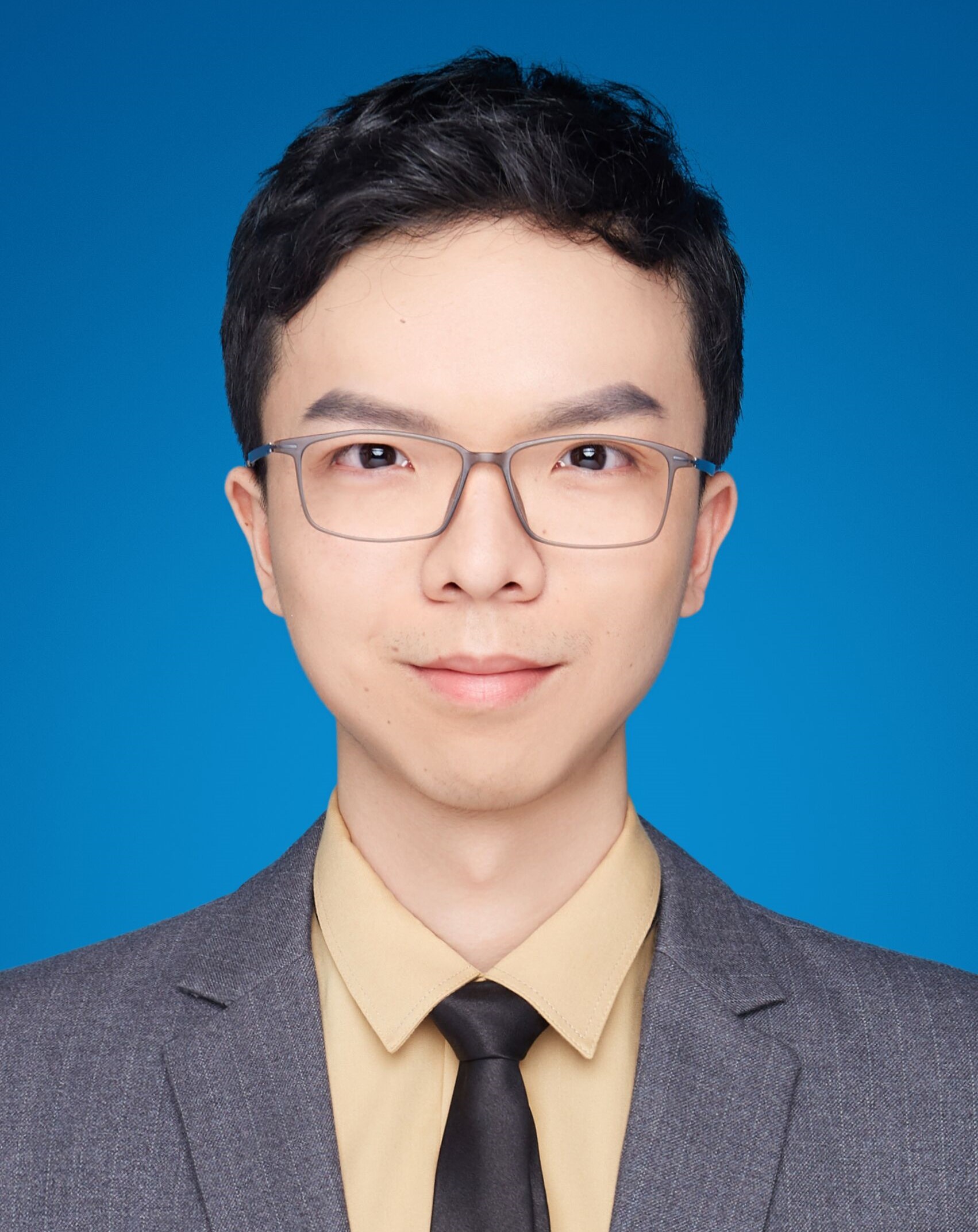}}]
{Sheng Zhang} is a first-year master student in the School of Informatics at Xiamen University. His research interests include data mining and recommender systems.
\end{IEEEbiography}

\vspace{-30pt}
\begin{IEEEbiography}[{\vspace{-10pt}\includegraphics[width=0.9in,height=1.15in,clip,keepaspectratio]{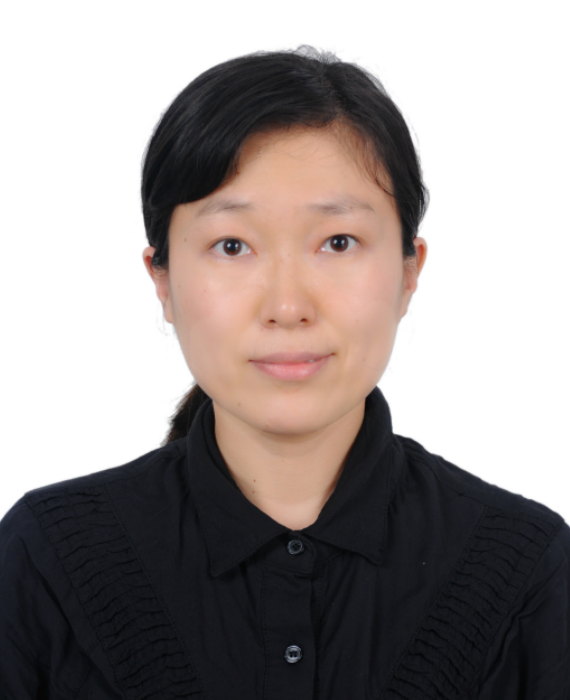}}]
{Min Gao} received the MS and PhD degrees in computer science from Chongqing University in 2005 and 2010 respectively. She is an associate professor at the School of Big Data \& Software Engineering, Chongqing University. Her research interests include recommender systems, Shilling Attack detection, and service computing.
\end{IEEEbiography}

\vspace{-45pt}
\begin{IEEEbiography}[{\includegraphics[width=1in,height=1.25in,clip,keepaspectratio]{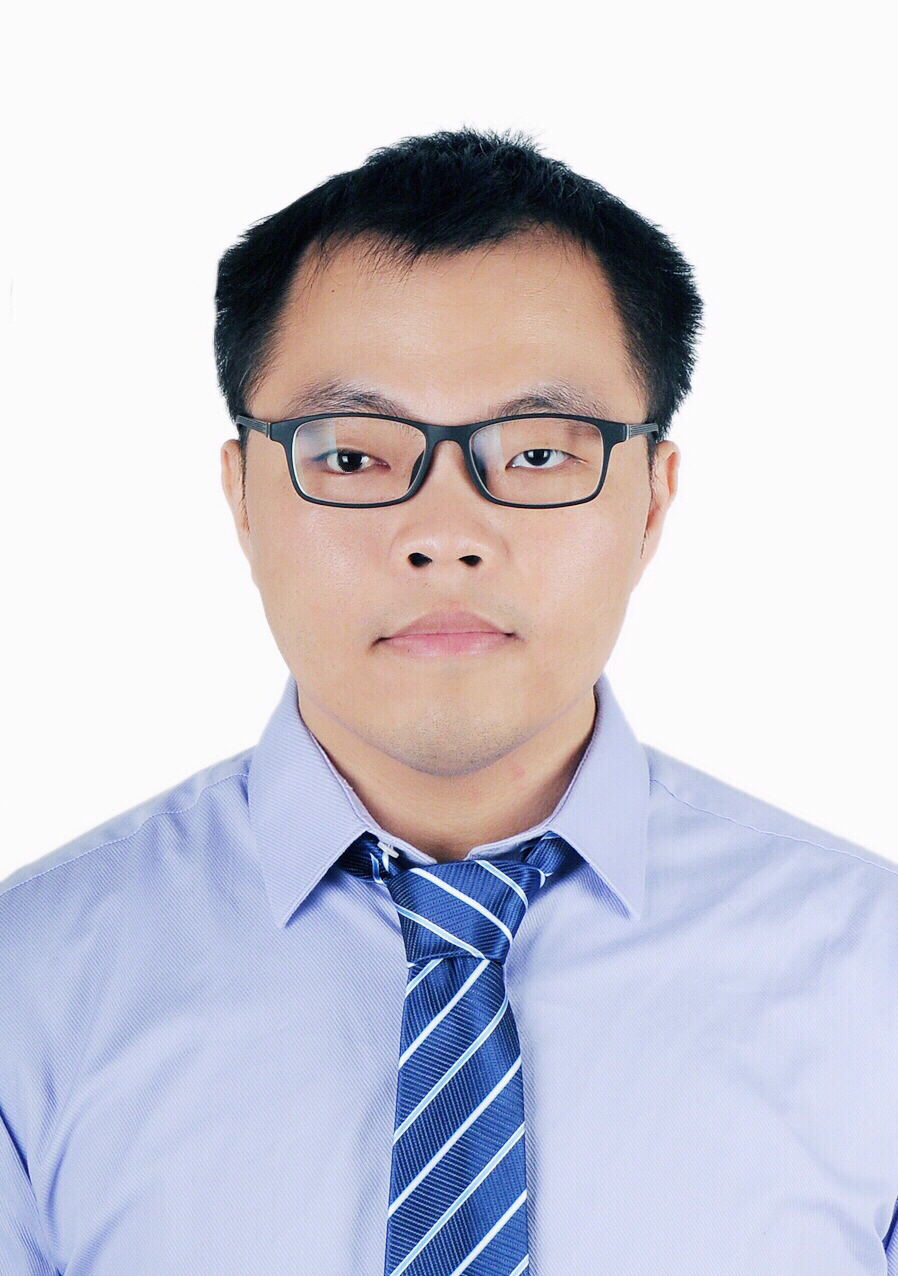}}]
{Hui Li} is currently an assistant professor in the School of Informatics, Xiamen University. His research interests include data mining and data management with applications in recommender systems and knowledge graph. He received his B.Eng. degree in Software Engineering from Central South University (2012), and his M.Phil. and Ph.D. degrees in Computer Science from University of Hong Kong (2015, 2018).
\end{IEEEbiography}

\end{document}

%% file: 1_intro.tex

\section{Introduction}
\label{sec:intro}
\IEEEPARstart{R}{ecommender} Systems (RS) have played a vital role
since the beginning of e-commerce~\cite{Aggarwal16}. The prevalence of RS in
industries (e.g., Amazon, Facebook and Netflix~\cite{ShiLH14}) has led to
growing interest in manipulating RS~\cite{DeldjooNM20}. On the one hand,
unscrupulous parties can gain illegal profits by attacking RS to mislead users and affect their decisions. On the other hand, studying how to
spoof RS gives insights into the defense against malicious attacks. 
We have seen various types of attacks against RS in the literature, including
Unorganized Malicious Attacks (i.e., several attackers individually attack RS
without an organizer)~\cite{Pang0TZ18}, Sybil Attacks (i.e., attacker illegally
infers a user's preference)~\cite{CalandrinoKNFS11}, etc. 

\begin{figure}[t]
\begin{center}
\includegraphics[width=0.95\columnwidth]{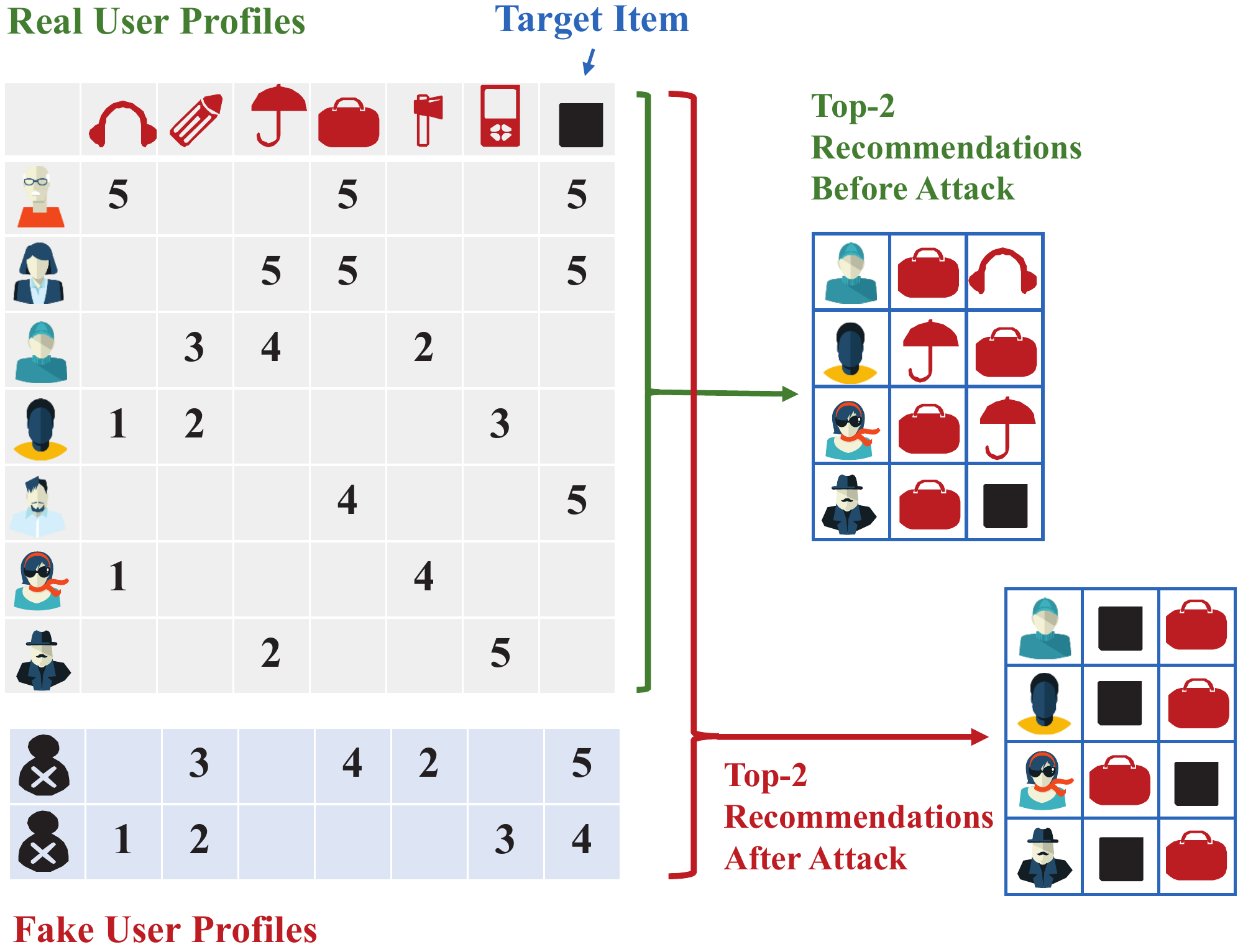}
\caption{An illustrative example of Shilling Attacks: promoting products by injecting fake user profiles. The value indicates the preference of a user on an item.}
\label{fig:example}
\end{center}
\end{figure}

This paper studies \emph{Shilling Attack}, which is one of the most subsistent
and profitable attacks against RS~\cite{GunesKBP14}. Shilling Attack is also called
as Data Poisoning Attack~\cite{LiWSV16,ChenL19} or Profile Injection Attack~\cite{BurkeMBW05}
in the literature. Researchers have successfully performed Shilling Attacks against real-world RS
such as YouTube, Google Search, Amazon and Yelp in
experiments~\cite{XingMDSFL13,YangGC17}. Large companies like Sony, Amazon and
eBay have also reported that they suffered from Shilling Attacks in
practice~\cite{LamR04}. In Shilling Attacks, an adversarial party performs an
attack by injecting a number of fake user profiles to hoax RS for improper
purposes~\cite{LamR04}. The goal is to either promote its own products (i.e.,
make the products recommended more often) or demote its competitors' products
(i.e., make the competitors recommended less often). Fig.~\ref{fig:example}
illustrates the former case: after injecting carefully crafted fake user
profiles, the target item appears in the top-2 recommendation list
provided by the victim recommendation model to some users. Shilling Attacks
fully utilize the core of RS: RS must allow users to interact with
the system by various operations such as giving ratings and browsing pages. This
way, RS can gain sufficient feedback from users to train their models and
provide recommendations. The open of RS to the data input from users makes
injecting fake user profiles to launch Shilling Attacks possible. Moreover, Shilling Attacks treat the
victim RS as a black box and only require the historical user-item
interaction data (e.g., ratings) which is typically accessible from real
users' public pages in the system. For instance, the pages of users containing
their historical data in local business RS platforms
Yelp\footnote{https://www.yelp.com} and
Dianping\footnote{http://www.dianping.com} are both open to the public.

Early Shilling Attack methods create injection profiles based on simple heuristics~\cite{GunesKBP14,LamR04,SandvigMB08}. For instance, Average Attack~\cite{LamR04} assigns the highest rating to the target item to be promoted and average ratings to a set of randomly sampled items. Recently, with the success of Adversarial Attacks~\cite{YuanHZL19} in image classification~\cite{SuVS19} and text classification~\cite{AlzantotSEHSC18}, a few works~\cite{abs-1809-08336,Christakopoulou19,Song20,Tang2020Revisiting} consider directly adopting an Adversarial Attack model for Shilling Attacks. 
Though we have witnessed a great success of Adversarial Attacks against many learning systems, existing Adversarial Attacks cannot be directly adopted for Shilling Attacks: 
\begin{enumerate}
\item \textbf{Attack goals are different.} Adversarial Attacks on text and image usually aim to trick the system to misclassify, while Shilling Attacks aim to misguide RS to rank the target item higher/lower. 
\item \textbf{Attack knowledge is different.} A prevalent strategy of Adversarial Attacks is to utilize the information of gradient decent in machine learning models to search undetectable perturbations~\cite{YuanHZL19}. However, in Shilling Attacks, though the data (e.g., rating matrix) of RS is generally available to all users (i.e., a user can see all other users' ratings) and thus exposed to attackers~\cite{SandvigMB08,GunesKBP14,LamR04}, the victim RS model is typically treated as a black box. 
\item \textbf{Data correlation is different.} Attacking many machine learning tasks can be achieved by manipulating one data sample (e.g., One-pixel Attack~\cite{SuVS19}). This is impractical for RS since recommendation for a user is typically made based on the information from multiple user-item pairs.  
\end{enumerate}

Besides, existing Shilling Attack methods suffer from the following limitations:
\begin{itemize}
\item\textbf{Low Attack Transferability.}
Simple heuristic based attacking methods are shown to be effective only on certain traditional collaborative filtering (CF) approaches. For example, Average~\cite{LamR04}, Bandwagon~\cite{Burke2005Limited} and Random~\cite{LamR04} Attacks are more effective against CF using user-based k-Nearest Neighbors, but do not work well against CF using item-based k-Nearest Neighbors~\cite{MobasherBBW07}. Since they are tailored for one victim RS model, their attack transferability on other victim RS models, especially on recently prevalent deep learning based RS models~\cite{ZhangYST19}, is doubtful.
\item\textbf{Low Attack Invisibility.}
Both simple heuristic based and most recent methods which directly use Adversarial
Attack models for Shilling Attacks lack the consideration of \emph{personalization}. The generated fake
user profiles do not have real-user behavior patterns in RS. Thus the attack can be
easily detected~\cite{Christakopoulou19,Lin2020Attacking}. 
Moreover, recent Adversarial Attack
models for Shilling Attacks against RS~\cite{Song20,Tang2020Revisiting} generate fake user profiles with implicit feedback,
i.e., profiles only have binary entries to indicate whether the
user has interacted with an item. To launch Shilling Attacks using these ``binary'' fake user profiles, attackers have to use bots to automatically trigger interactions (e.g., frequently browse the page of an item) and mislead the system to regard the frequent operations as an indication of user preferences. Such frequent operations may arouse the suspicion of the detector. 
\end{itemize}

In this paper, we propose \ours (\textbf{Le}arning to \textbf{G}enerate Fake \textbf{U}ser \textbf{P}rofiles), which extends our previous work AUSH~\cite{Lin2020Attacking} for Shilling Attacks, and is built upon the Generative Adversarial Network (GAN)~\cite{GoodfellowPMXWOCB14}.
The novelty of \ours lies in the following designs to enhance its attack transferability and invisibility, which are different compared to AUSH and other recent Shilling Attack methods:
\begin{enumerate}
\item \textbf{Generator: a neural network module to produce discrete ratings.}
The generated fake user profiles of \ours include discrete ratings. 
As giving Likert-scale ratings is the most common way that real RS provide to users to give feedback and get involved in the recommendation procedure, the fake user profiles are easy to inject into the system (e.g., type in ratings on a limited number of items manually), but difficult for the detector to discover.
\ours adopts a discretization layer in the generator which produces learnable thresholds to generate ratings with personalized user behavior patterns. To avoid the problem that discretization will encounter unpropagated gradients in training, we train \ours through an approximation function.
Since the error of the discretization will be eliminated during optimization, the generated fake user profiles are more accurate and powerful in Shilling Attacks. This way, both attack transferability (i.e., attack performance) and invisibility of \ours are enhanced.

\item \textbf{Generator Optimization: indirect and direct generation losses.}
The generator in a standard GAN is usually optimized through the classification task in the discriminator.
Towards better attack performance, a \emph{generation loss} can be associated solely with the generator. In AUSH, the generation loss is \emph{indirect}, i.e., it is related to the reconstruction of some user data.
Since the generator acts like an ``attacker'', the generation loss can also be \emph{direct}, i.e., to measure how well the victim RS model is attacked.
Inspired by Tang et al.~\cite{Tang2020Revisiting}, we design a generation loss to be estimated on a surrogate model for \ours. We also propose a two-level optimization for the generation loss.
Thus, less knowledge is demanded regarding the victim RS model, black-box attack is allowed, and higher attack transferability can be achieved.

\item \textbf{Discriminator.}
In GANs, the generator and the discriminator strike to enhance each other in a minimax competition.
Specifically, in Shilling Attacks, by training the generator to ``beat" the discriminator, the generated fake user profiles are less detectable and more powerful in the attack.
We provide a design of the discriminator for \ours, which acts like a ``defender'' and identifies fake user profiles based on both explicit and implicit feedback, so that both attack transferability and attack invisibility can be improved.
\end{enumerate}

We have conducted comprehensive experiments to verify the effectiveness and
undetectability of \ours. We show that \ours is effective against a wide range
of victim RS models, including both classic and modern deep learning based
recommendation algorithms, on a variety of benchmark RS datasets. We also show
that \ours is virtually undetectable by the state-of-the-art attack detection
method. We demonstrate, by visualizations, that the fake user profiles
produced by \ours follow a similar distribution as real users and maintain
diversity (i.e., personalization). We hope that, \ours, as a new Shilling
Attack method, can benefit the study of secure and robust RS. 

The rest of the paper is organized as follows: Sec.~\ref{sec:related} 
surveys the related work. Sec.~\ref{sec:bg} introduces the background of
Shilling Attacks against RS and the framework design of \ours.
Sec~\ref{sec:implementation} and Sec.~\ref{sec:learning} describe in detail
\ours's model architecture and learning algorithm, respectively. In
Sec.~\ref{sec:experiment}, we compare \ours with other state-of-the-art
Shilling Attack methods to verify its effectiveness and undetectability. Finally,
Sec.~\ref{sec:con} concludes our work and points out future research
directions.

%% file: 2_related_work.tex

\section{Related Work}
\label{sec:related} 

\subsection{Recommender Systems (RS)}

RS can help overcome the information overload problem. The research on RS has a long history. Early work on RS is either heuristic-based or factorization-based~\cite{abs-2104-13030}.

Recently, the great success of deep learning has significantly advanced the development of RS.  
Following are some representative deep neural networks adopted in RS and their examples:
(1) Convolutional Neural Network (CNN) can capture local and global representation from heterogeneous data~\cite{TuanP17}. One recent work RecSeats~\cite{MoinsAB20} adopts CNN to capture higher order interactions between features of available seats in seat recommendation. 
(2) Multilayer Perceptron (MLP) can easily model the nonlinear interactions between users and items even when the system has much noise data. Recently, Zhou et al.~\cite{abs-2202-13556} enhance MLP with filtering algorithms from signal processing that attenuates the noise in the frequency domain. Experimental results show that their method can significantly improve recommendation quality.
(3) Recurrent Neural Network (RNN) can help RS model sequential behaviors better. For instance, Zhang et al.~\cite{zhang2022deep} recently adopt RNN in their proposed deep dynamic interest learning that captures the local/global sessions within sequences for CTR prediction and Xia et al.~\cite{XiaWDZCC19} leverage RNN for modeling reviews in RS.
(4) Graph Neural Network (GNN) can model structural information in graph data that is prevalent in RS. 
Huang et al.~\cite{HuangDDYFW021} study how to use GNN together with negative sampling to provide recommendations on user-item graphs.
Huang et al.~\cite{HuangXXDXLBXLY21} model both the high order user- and item-wise relation encoding in GNN for social recommendation.

Due to the page limit, we only illustrate recent, representative RS. Readers can refer to related surveys~\cite{ZhangYST19,abs-2104-13030} for more related work on RS.

\subsection{Adversarial Attacks} 
Investigating the security of machine learning based systems is a continuing
concern within the machine learning community.  Research has shown that,
crafted adversarial examples~\cite{YuanHZL19}, which may be imperceptible
to the human eye, can lead to unexpected mistakes of machine learning based
systems. Adversarial Attacks, which are launched by adversaries to leverage
vulnerabilities of the system, has been studied in many text and image based
learning systems~\cite{Zhang20,YuanHZL19}.

Adversarial examples in conventional machine learning models have been
discussed since decades ago~\cite{YuanHZL19}. Dalvi et al.~\cite{DalviDMSV04}
find that manipulating input data may affect the prediction results of
classification algorithms. Biggio et al.~\cite{BiggioCMNSLGR13} design a
gradient-based approach to generate adversarial examples against SVM. Barreno
et al.~\cite{BarrenoNSJT06,BarrenoNJT10} formally investigate the security of
conventional machine learning methods under Adversarial Attacks. Roli et
al.~\cite{RoliBF13} discuss several defense strategies against Adversarial
Attacks to improve the security of machine learning algorithms. In addition to
conventional machine learning, recent studies have reported that deep learning
techniques are also vulnerable to Adversarial
Attacks~\cite{YuanHZL19}.

Even though Adversarial Attack methods are able to affect many learning applications, we cannot directly apply these methods to Shilling Attacks as explained in Sec.~\ref{sec:intro}.

\subsection{Generative Adversarial Network}
Generative Adversarial Network (GAN)~\cite{GoodfellowPMXWOCB14} performs adversarial learning
between a generator and a discriminator, which can be implemented with any form of differentiable system that
maps data from one space to the other. The generator
tries to capture the real data distribution and generates real-like data,
while the discriminator is responsible for discriminating the generated
and original data. GAN plays a minimax game and the
optimization terminates at a saddle point that is a minimum with respect to
the generator and a maximum with respect to the discriminator (i.e., Nash
equilibrium). 

As GAN overcomes the limitations of previous generative
models~\cite{HongHYY19}, it has been prevalently deployed in  applications
that generate text~\cite{YuZWY17}, images~\cite{GoodfellowPMXWOCB14},
recommendations~\cite{WangNL2019} or many other types of
data~\cite{HongHYY19}. To improve original GAN, DCGAN~\cite{RadfordMC15}
adopts the CNN architecture, and Wasserstein GAN~\cite{ArjovskyCB17} leverages
Earth Mover distance. There also exists a direction of GAN research which
utilizes GAN to generate adversarial examples. For instance, Zhao et
al.~\cite{ZhaoDS18} propose to search the representation space of input data
under the setting of GAN in order to generate more natural adversarial
examples. Xiao et al.~\cite{XiaoLZHLS18} design AdvGAN which can attack
black-box models by training a distilled model. 

\subsection{Shilling Attacks against RS} 

O'Mahony et al.~\cite{OMahonyHS05,OMahonyHKS04} firstly study the robustness of
user-based CF by injecting some fake users. They
also provide a theoretical analysis of the attack by viewing injected ratings
as noises. Lam and Riedl~\cite{LamR04}, Burke et al.~\cite{BurkeMBW05,Burke2005Limited}, Mobasher et al.~\cite{MobasherBBW07} 
further study the influence of some low-knowledge attack approaches to promote an item (e.g.,  
Random, Average, Bandwagon and Segment Attacks) and demote an item 
(e.g., Love/Hate Attacks and Reverse Bandwagon Attacks) on CF methods.
Assuming more knowledge and cost are available, Wilson and Seminario~\cite{WilsonS13}, and Seminario and Wilson~\cite{SeminarioW14b} design 
Power User/Item Attacks which leverage most influential users/items to
shill RS, Fang et al.~\cite{FangYGL18} study how to spoof graph based RS
models, and Li et al.~\cite{LiWSV16} present near-optimal Data Poisoning
Attacks for factorization-based RS. Xing et al.~\cite{XingMDSFL13} and Yang et
al.~\cite{YangGC17} conduct experiments on attacking real-world
RS (e.g., YouTube, Google Search, Amazon and Yelp), and show that manipulating
RS is possible in practice. 

The success of Adversarial Attacks have inspired the study of Shilling
Attacks. Christakopoulou and Banerjee~\cite{abs-1809-08336,Christakopoulou19}
employ DCGAN~\cite{RadfordMC15} to generate fake user profiles used in
Shilling Attacks. However, directly adopting existing GANs will not provide
satisfactory results in Shilling Attacks as shown in our experiments. Tang et
al.~\cite{Tang2020Revisiting} leverage a surrogate victim model to estimate
the attack performance for black-box attacks. 
Song et al.~\cite{Song20} design an effective Shilling Attack framework
based on reinforcement learning. But their method requires that the feedback
from RS is periodically available, which is impractical for most RS. Our
previous work AUSH~\cite{Lin2020Attacking} is tailored for RS by considering
attack cost and designing a specific generation loss. But AUSH applies simple
rounding to generate discrete ratings, and its generation loss is indirect
and only related to the reconstruction of some user data, which may affect the
results of Shilling Attacks.

%% file: 3_model.tex
\section{Background and Overview of \ours}
\label{sec:bg}

\subsection{Terminology}
\label{sec:background}

First, we introduce the terminology of Shilling Attacks:

\begin{itemize}
\item \textbf{Attack goal.} The goal of an adversarial party in Shilling Attacks could be complex. In this paper, we mainly consider targeted attack, i.e., one item must be influenced. This is the most common case for Shilling Attacks against RS, because retail companies and producers have a strong intention to increase sales in the fierce business competition.
	\begin{itemize}
		\item \textbf{Push Attacks} indicate that one or several target items must be promoted, i.e., the target items must be recommended by the victim RS model more than they were before the attack.
		\item \textbf{Nuke Attacks} indicate that one or several target items must be demoted. Although we use Push Attacks to demonstrate the model design throughout this paper, it is convenient to apply the techniques to Nuke Attacks by reversing the goal setting.
	\end{itemize}
\item \textbf{Attack budget.} Attacking RS is costly. When designing a practical Shilling Attack method against RS, we have to take into account the attack budgets from two perspectives:
	\begin{itemize}
	\item \textbf{Attack size} is the number of fake user profiles. The larger the attack size is, the more effective and expensive the attack could be.
	\item \textbf{Profile size} is the number of non-zero ratings in one fake user profile. Some recent works~\cite{Tang2020Revisiting} do not consider profile size. However, we believe that constraining the profile size is necessary. In some E-commerce platforms, injecting fake ratings is impossible without real purchase. Thus, it will be too expensive to generate fake users with unlimited non-zero ratings.
	\end{itemize}
\item \textbf{Attack knowledge.} How much knowledge is accessible for the attacker is a critical factor in designing Shilling Attack methods. In general, the most desirable knowledge is related to the user feedback and the victim RS model:
	\begin{itemize}
		\item \textbf{User feedback} is the dataset used to train the victim RS model. The attacker can have full or partial knowledge about the user feedback. In this paper, we assume that the attacker has full knowledge of the user feedback, i.e., the attacker knows who rates what and the exact rating values. This is a reasonable setting and is commonly adopted in literature~\cite{SandvigMB08,GunesKBP14,LamR04,Tang2020Revisiting}, because user-item ratings in RS are generally available to all users (e.g., a user can see all other users' ratings) and thus exposed to attackers.
		\item \textbf{Victim RS model} is used to deliver recommendations and it is the target of Shilling Attacks. Some existing works assume that white-box attack is possible, i.e., the attacker has different degrees of knowledge about the victim model, e.g., the model type~\cite{FangYGL18} and model parameters~\cite{Mei2015Using}. In this paper, we assume that the victim model is unseen to the attacker (i.e., a black box). This is because in many real productive RS, the model is so complex that acquiring extraordinary knowledge about the model parameters is impossible. Furthermore, real RS usually employ ensemble models and update their models frequently~\cite{Aggarwal16}. Therefore, the attacker can not have accurate knowledge about the model type.
	\end{itemize}
	\item \textbf{Injected fake user profiles.} Although \ours does not explicitly classify items in a user profile, we follow the terminology used in the literature~\cite{GunesKBP14} and divide the items in a fake user profile into four parts so that our descriptions are consistent with previous works:
	\begin{itemize}
		\item \textbf{Target item} indicates the item that the attacker wants to fulfill his/her malicious purpose.
		\item \textbf{Filler items} are the items which have non-zero ratings in the injected fake user profiles. The filler items in each fake user profile are usually different.
		\item \textbf{Unrated items} are the items that have not been assigned with any ratings in the injected fake user profiles. 
		\item \textbf{Selected items} are several human-selected items for a special treatment. Not all attack models consider selected items. One possible reason to use selected items is to influence \emph{in-segment users}, i.e., users that have shown preferences on selected items. Details can be found in Sec.~\ref{sec:secondary}.
	\end{itemize}
\end{itemize}

\begin{figure*}[t]
\begin{center}
\includegraphics[width=1\textwidth]{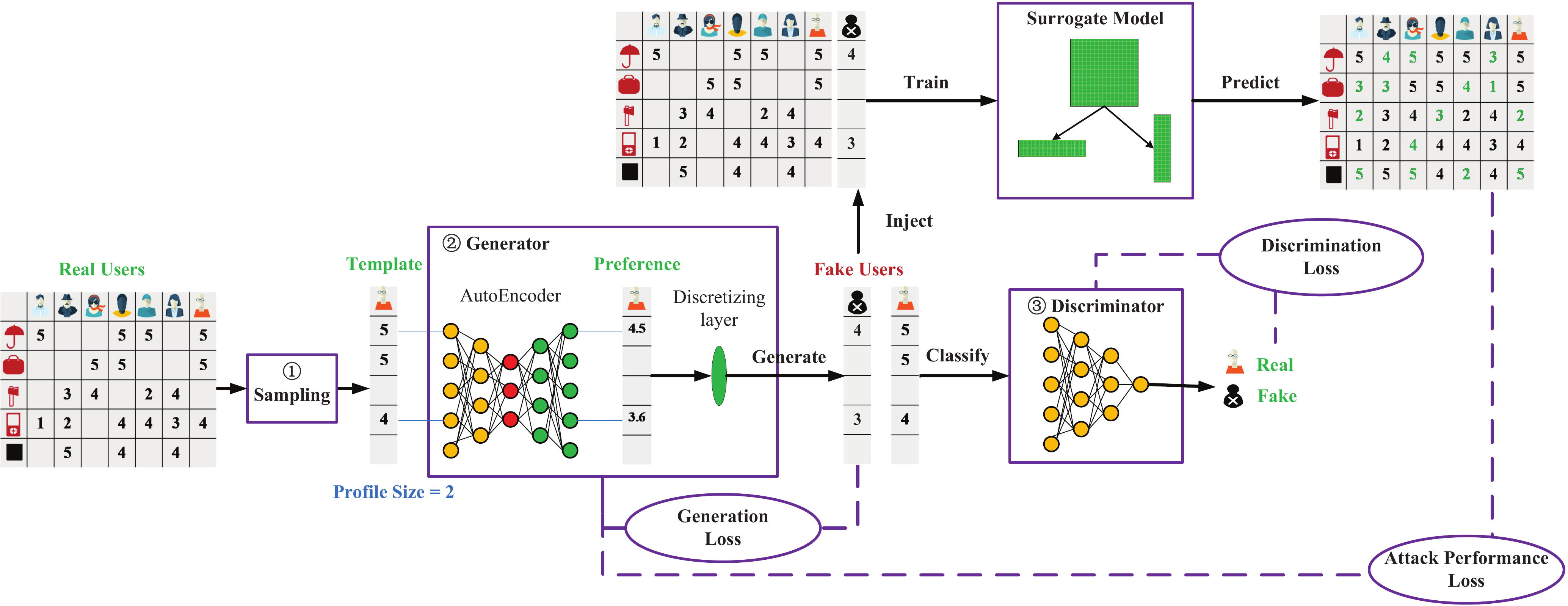}
\caption{Overview of \ours. \ours consists of three parts: (1) A sampling step that samples real user profiles as ``templates''; (2) A generator that generates fake users; (3) A discriminator that distinguishes real user profiles and fake user profiles to boost the generator's ability to generate undetectable fake user profiles.}
\label{fig:framework}
\end{center}
\end{figure*}

\begin{table}[t]
\caption{Notations for \ours.}
\label{tab:notations}
\centering
\scalebox{0.9}{
\begin{tabular}{|l|p{5cm}|}
\hline
\multicolumn{1}{|c|}{Notation}                                                 & \multicolumn{1}{c|}{Definition} \\                                                                                                                  \hline
$\users, \fakeusers, \items,\select$ & Sets of real users, fake users, items, and selected items. \\ \hline
 $\inputrating \in \real^{|\users |\times |\items|}$ & Input rating matrix of real users and items.                                                                                            \\ \hline
$G_\Gparam(\inputrating^{(in)})\text{ or }\outputrating \in \real^{|\fakeusers| \times |\items|}$   & Injected fake user rating matrix.                                                                                 \\ \hline
$\sample \in \real^{A \times |\items|}$   & Sampled user rating matrix used for ``templates".                                                                                 \\ \hline
$\predict \in \real^{*\times *}$ & RS's predictions on some users and items. \\\hline
$\inputrating_{u,:}$ & A row of $\inputrating$ represents a user $u$. \\\hline
$\inputrating_{:,i}$ & A column of $\inputrating$ represents an item $i$. \\\hline
$\items_u=\{i \in \items: \inputrating_{u,i}\neq 0\}$ & The item set that user $u$ has rated. \\\hline
 $\users_i=\{u\in \users: \inputrating_{u,i}\neq 0\}$ & The set of users who have rated item $i$. \\\hline
$\outputrating=\Gen$                       & The generator producing fake user profiles based on ``templates'', parameterized by $\Gparam$. \\ \hline
$D_{\Dparam}(\outputrating)$     & The discriminator parameterized by $\Dparam$.      \\ \hline
$O_{\Sparam}\big(\inputrating,\outputrating\big)$     & The surrogate victim RS model parameterized by $\Sparam$.    \\
\hline
\end{tabular}
}
\end{table}

\subsection{Notations}
Throughout this paper, we use lower-case letters for indices, capital letters for scalars, boldface lower-case letters for vectors, boldface capital letters for matrices, and calligraphic letters for sets. The notations used for \ours are illustrated in Tab.~\ref{tab:notations}.
We use $\inputrating \in \real^{|\users |\times |\items|}$ to denote the real rating matrix in RS, where $\users$ and $\items$ represent the user universe and the item universe, respectively.
Each row of $\inputrating$, i.e., $\inputrating_{u,:}$ represents the ratings given by user $u$. Each column of $\inputrating$, i.e., $\inputrating_{:,i}$ represents the ratings assigned to item $i$.
$\items_u=\{i \in \items: \inputrating_{u,i}\neq 0\}$ indicates the set of items that have been rated by user $u$.
Similarly,  $\users_i=\{u\in \users: \inputrating_{u,i}\neq 0\}$ denotes the set of users that have rated item $i$.
The attack budget is given by $A$ and $P$, i.e., the attack size and profile size.
\ours takes $\inputrating$ as the input and generates the fake user profiles $\outputrating$, where $\outputrating \in \real^{|\fakeusers| \times |\items|}$.
Since the attack size is $A$, $|\fakeusers|=A$. Given the profile size $P$, there are exactly $P$ non-zero entries in $\outputrating_{v,:}$, $\forall v\in \fakeusers$.

\subsection{Framework Overview}
\label{sec:overview}
Inspired by the success of GAN based adversarial learning in text~\cite{YuZWY17} and image~\cite{GoodfellowPMXWOCB14} generation, the design of \ours follows a GAN framework.
Fig.~\ref{fig:framework} gives an overview of our pipeline, which consists of the following parts:

\begin{enumerate}
\item\textbf{Sampling.}
Making up fake user profiles from scratch is risky as the generated profiles may not have real-user patterns.
Following our previous work AUSH~\cite{Lin2020Attacking}, we use sampled real user profiles as ``templates'' and learn real user patterns from them. Using ``templates'' instead of all the real users for learning real user patterns is computationally efficient and makes it easy for \ours to focus on capturing patterns of real users with rich information which leads to high-quality fake profiles. The knowledge of templates is accessible in practice and does not exceed the requirements of recent sophisticated attack methods~\cite{LiWSV16,FangYGL18}, as we have shown in Sec.~\ref{sec:background}.
The sampling component in \ours gives the generator the opportunity to learn from real user behaviors and retain the preference diversity of the community (i.e., personalization).
Specially, the sampling component contains two steps. In the first step, a batch of real users are chosen as ``templates''. The second step is optional. When a fixed profile size is required, in the second step, filler items are sampled from the rated items of the corresponding ``template''.
\item\textbf{Generator.}
Generator takes as input the sampled user-item rating sub-matrix (i.e., ``templates'') and captures the latent association between items and users.
It outputs the injected fake user profiles in the form of discrete ratings.
To boost the attack performance, the parameters of generator are learned by minimizing a \emph{generation loss}, which measures how much the victim RS model will be affected after the attack.
As we are performing black-box attacks, the details of the victim recommendation model is unknown. Thus, the generation loss is measured via a surrogate victim model in \ours.
\item\textbf{Discriminator.}
The discriminator of \ours is fed with the output of the generator.
It distinguishes real user profiles and fake user profiles.
Optimizing the discriminator, through a \emph{discrimination loss} which measures the accuracy of the classification of real/fake user profiles, boosts the generator's ability to generate undetectable fake user profiles.
\end{enumerate}

The design of \ours is general and there are various possible implementations for the generator and the discriminator. We provide the details of one implementation in Sec.~\ref{sec:implementation}.

%% file: 4_imp.tex
\section{Model}
\label{sec:implementation}

In this section, we will elaborate on one implementation for \ours to accomplish Shilling Attacks.

\subsection{Sampling}
\label{sec:sampling}
As illustrated in Sec.~\ref{sec:overview}, \ours contains a two-step sampling component:

\begin{enumerate}
\item\textbf{User Sampling}. The first step is to sample users $u \sim \users$ from the existing users to construct ``templates". The result is a sub-matrix of $\inputrating$, i.e., $\sample \in \real^{|\mathcal{U'}| \times |\items|}$.
Given the attack size $A$, we have $|\mathcal{U'}|=A$.
Each ``template'' is sampled randomly from real users.
\item\textbf{Filler Item Sampling}.
Given the profile size $P$, \ours uses the second step to select items in each ``template''. We follow the setting of AUSH~\cite{Lin2020Attacking} and randomly keep $P$ items if the sampled user (i.e., ``template'') has interacted with more than $P$ items. If the sampled user has interacted with less than $P$ items, we only use these items in this ``template''. 
\end{enumerate}

\subsection{Generator}
\label{sec:generator}

The generator aims to capture the real user preferences from ``templates'' in order to construct the fake user profiles in the form of discrete ratings for Shilling Attacks. There are various possible model
structures for the generator. Without loss of generality, we divide the
generator of \ours into two major components. One is used to generate user
preferences (i.e., numerical values in the latent space) from ``templates'', the other is to transform the user preferences to discrete ratings:

\vspace{5pt}
\noindent\textbf{Generating user preference.}
To infer user preferences in $\sample$, a function, \emph{preference learner}, that yields numerical output is required. Suppose the preference learner is denoted by $\Learner$, with parameters $\Lparam=\{\omega\}$. The simplest preference learner is to directly output its parameters:
\begin{equation}
\label{equ:simple}
\Learner_{v,i}=\omega_i.
\end{equation}

We can also adopt a more complex neural network module to extract user preferences from ``templates''. One attractive feature of using such a module is that it can capture non-linear user-item associations in the ``templates''. Since most victim RS models deliver recommendations based on user-item correlations, adopting a complex preference learner can be helpful in shilling the victim RS model. To be specific, we empirically find that the following Autoencoder (AE) based design works well for \ours:
\begin{equation}
\begin{aligned}
\label{equ:AE}
\Learner_{v}&=f_{decode}\big(f_{encode}(\sample_{v,:})\big) * 2.5 + 2.5,\\
f_{encode}(\mathbf{X})&=ReLU(\mathbf{W}_e\mathbf{X}+\mathbf{b}_e),\\
f_{decode}(\mathbf{X})&=Tanh(\mathbf{W}_d\mathbf{X}+\mathbf{b}_d),\\
\end{aligned}
\end{equation}
where $\mathbf{W}$ and $\mathbf{b}$ are learnable weight matrices and bias vectors, respectively. 
In Eq.~\ref{equ:AE}, $f_{\mathit{encode}}(\sample_{v,:})$ is a Multilayer Perceptron (MLP) which transforms the ratings $\sample_{v,:}$ into a low dimensional latent representation, i.e., user preferences. $f_{\mathit{decode}}(\cdot)$ is another MLP that decodes the latent representation back to a rating vector. Note that, in the implementation, each template has at most $P$ non-zero entries, i.e., $\|\sample_{v,:}\|_0 \leq P$ and only the non-zero entries are involved in computing. To ensure that fake user profiles will have exactly $P$ non-zero ratings, we operate on the corresponding entries of the output, hence we have $\|\Learner_{v}\|_0\leq P$. Moreover, we normalize the output of the preference learner via first multiplying it by 2.5 and then adding 2.5 so that $0<\Learner_{v}<1$.

\vspace{5pt}
\noindent\textbf{Discretizing user preference.}
Giving discrete ratings (e.g., 5-star scale ratings) to items is the common way that RS provide to users to interact with the system and get involved in the recommendation procedure.
Thus, the numerical output of AE in \ours, which indicates the preference strength of each fake user on each filler item, is designed to be discretized as 5-star ratings.
Our previous work AUSH~\cite{Lin2020Attacking} adopts a simple rounding method to project numerical preference to discrete ratings. The rounding method in AUSH divides the range into five segments:
\begin{equation}\label{equ:rounding}
\outputrating_{v,i}=\begin{cases}
1, & 0 < \Learner_{v,i}\leq 0.2 \\
2, & 0.2 < \Learner_{v,i}\leq 0.4 \\
3, & 0.4 < \Learner_{v,i}\leq 0.6 \\
4, & 0.6 < \Learner_{v,i}\leq 0.8 \\
5, & 0.8 < \Learner_{v,i} < 1.  \\
\end{cases}
\end{equation}

However, the rounding method in AUSH is intuitively sub-optimal:
\begin{enumerate}
\item It can not discover subtle differences among users as the rounding method is identical to all users and/or items. For example, suppose the estimated preference is 0.85, a normal user may give a $5$-star rating, while a picky user might give a $4$-star rating. Rounding will output $5$-star for all users, which is not feasible.
\item It will incur unpredictable errors that influence attack performance. Since  rounding is performed when the optimization of AUSH is completed, errors introduced by discretizing the preference are not accumulated in the learning objective. Suppose that the estimated preference from AUSH is $4.8$, which has minimized the generation loss of AUSH. Then, discretizing it to $5$ may not conform to the direction of optimization for the generation loss, which will affect the attack performance.
\end{enumerate}

To address above issues in AUSH, we design a final discretization layer in the generator of \ours to discretize preference so that the discretization process becomes part of the optimization and discretization errors can be reduced:
\begin{equation}
\label{equ:generator}
G_\Gparam(\sample)_{v,i}=\sum_{k=1}^{5} H\Big(\Learner_{v,i}, \sum_{k'=1}^{k} \tau_{v,k'}\Big),
\end{equation}
where $\tau$ is a learnable parameter, $\forall v\in \fakeusers, 0 \leq \tau_{v,k}\leq 1$. $\Gparam=\{\mathbf{\tau},\mathbf{W},\mathbf{b}\}$ is the set of parameters of AE and the final discretization layer. $H(x,\tau), 0<x<1$ is the Heaviside step function defined as follows: 
\begin{equation}
\label{eq:Heaviside}
H(x,\tau) = \begin{cases}
1, & x>\tau \\
0, & x\leq \tau. \\
\end{cases}
\end{equation}

Essentially, the final discretization layer learns five value segments for each user $v$:
\begin{equation}
\outputrating_{v,i}=\begin{cases}
1, & 0 < \Learner_{v,i}\leq \tau_{v,1} \\
2, & \tau_{v,1} < \Learner_{v,i}\leq \sum_{k=1}^2 \tau_{v,k} \\
3, & \tau_{v,1}+\tau_{v,2} < \Learner_{v,i}\leq  \sum_{k=1}^3 \tau_{v,k} \\
4, & \sum_{k=1}^3 \tau_{v,k} < \Learner_{v,i}\leq \sum_{k=1}^4 \tau_{v,k} \\
5, & \sum_{k=1}^{4} \tau_{v,k} < \Learner_{v,i} < \sum_{k=1}^{5} \tau_{v,k}.   \\
\end{cases}
\end{equation}

We can attach a generation loss with the generator to enhance the quality of generated fake user profiles. It is worth noting that, even without the generation loss, the generator can still get rewarded/penalized through the discrimination loss introduced in Sec.~\ref{sec:dis}.
Generation loss can be \emph{indirect}. For example, AUSH employs a \emph{reconstruction loss} which measures how well the generated ratings recover the real ratings:
\begin{equation}
\label{equ:reconstruction}
\gloss=\sum_{v\in\users} \sum_{\sample_{v,j}\neq 0} \big(\outputrating_{v,j}-\sample_{v,j}\big)^2.
\end{equation}

Eq.~\ref{equ:reconstruction} does not require any prior knowledge of the victim RS model and it is irrelevant to the attack. The design goal of the reconstruction loss is to help the fake user profiles retain the real user behavior patterns so that they can be difficult to detect.

We can also \emph{directly} maximize the attack performance via the generation loss. We use Push Attacks as an example, i.e., how much the target item $t$ will be promoted:
\begin{equation}
\label{equ:gloss}
\gloss=-\sum_{u\in\users}\log \frac{\exp{\predict_{u,t}}}{\sum_{j\in \items} \exp{\predict_{u,j}}},
\end{equation}
where $\predict_{u,i}$ is the rating predicted by the victim model for user $u$ on item $i$.
If the target item $t$ receive a higher predicted rating by the victim RS model than other items after the attack, then the direct loss in Eq.~\ref{equ:gloss} will be minimized. Note that Nuke Attacks can be achieved similarly via maximizing Eq.~\ref{equ:gloss}.

However, we can not obtain the predicted ratings $\predict$ by the victim RS model in the setting of the black-box Shilling Attacks.
Instead, we use the output of a surrogate RS model to represent $\predict$, with the assumption that the knowledge for attacking a well-trained surrogate RS model can be transferred to attacking other RS models (i.e., real victim RS models)~\cite{Tang2020Revisiting}. Therefore, the predicted ratings $\predict$ can be obtained via:
\begin{equation}
\begin{aligned}
\predict&=S_{\Sparam}\big(\inputrating^{*}\big),\\
\inputrating^{*}&=concate\big(\inputrating,G_\Gparam(\inputrating^{(in)})\big),
\end{aligned}
\end{equation}
where $S_{\Sparam}$ indicates a surrogate RS model with parameters $\Sparam$, and $\inputrating^{*}$ indicates the concatenation of two matrices $\inputrating$ and $G_\Gparam(\inputrating^{(in)})$. 
Note that, the surrogate model is trained on both real user feedback and fake user profiles to achieve a minimal recommendation loss $\rsloss$:
\begin{equation}
\Sparam=\arg \min_{\Sparam} \rsloss\big(\inputrating^{*}, S_{\Sparam}(\inputrating^{*})\big).
\end{equation}
We use the original design of the recommendation loss $\rsloss$ in the surrogate RS. But we calculate $\rsloss$ on $\inputrating^{*}$ instead of $\inputrating$. 

In summary, we have the following objective for the generator:
\begin{equation}
\begin{aligned}
\label{equ:gobjective}
\min_\Gparam\,\,\, &-\sum_{u\in\users}\log \frac{\exp{S_{\Sparam}\big(\inputrating^{*}\big)_{u,t}}}{\sum_{j\in \items} \exp{S_{\Sparam}\big(\inputrating^{*}\big)_{u,j}}}\\
s.t., \Sparam &=\arg \min \rsloss\big(\inputrating^{*}, S_{\Sparam}(\inputrating^{*})\big).
\end{aligned}
\end{equation}

The choice of the surrogate model is important to ensure attack transferability. Following Tang et al.~\cite{Tang2020Revisiting}, we use Weighted Regularized Matrix Factorization (WRMF) as the default surrogate model due to its efficiency and effectiveness on real data~\cite{Aggarwal16,ShiLH14}. But we also explore other surrogate models in our experiments in Sec.~\ref{sec:surrogate}. In WRMF, the recommendation loss $\rsloss$ is defined as the weighted aggregation of the differences between predictions and observed ratings in $\inputrating^{*}$:
\begin{equation}
\begin{aligned}
\label{equ:rloss}
&\rsloss\big(\inputrating^{*}, S_{\Sparam}(\inputrating^{*})\big)\\
=&\sum_{u\in\users, i\in\items} w_{u,i} \big(\inputrating^{*}_{u,i}-\mathbf{P}_u^T\mathbf{Q}_i\big)^2+\lambda(\|\mathbf{P}\|^2 + \|\mathbf{Q}\|^2),
\end{aligned}
\end{equation}
where $w_{u,v}$ is the instance weight for the observed rating $\inputrating_{u,i}\neq 0$ and the missing rating $\inputrating_{u,i}=0$. $\Sparam=\{\mathbf{P},\mathbf{Q}\}$ are model parameters of the surrogate model, and $\lambda$ is the hyper-parameter to control model complexity. 

\subsection{Discriminator}
\label{sec:dis}
The discriminator $D$ attempts to correctly distinguish fake user profiles and real user profiles.
It takes the fake user profiles generated by the generator or the real user profiles, and outputs the probabilities that the inputs are real.
We use a MLP as our discriminator.
For simplicity, we illustrate the discriminator in Eq.~\ref{equ:D} with the fake user profiles $\outputrating$:
\begin{equation}
\label{equ:D}
D_{\Dparam}(\outputrating) = \sigma\big(\mathbf{W}_d(\outputrating)+\mathbf{b}_d\big),
\end{equation}
where $\Dparam=\{\mathbf{W}_d,\mathbf{b}_d\}$ 
are learnable weight matrices and bias vector, and $\sigma$ indicates the sigmoid function. Note that we use both explicit feedback (non-zero ratings) and implicit feedback (denoted as zeros) in the user profiles for Eq.~\ref{equ:D}.

The goal of the discriminator is essentially a binary classification task. We use the cross-entropy loss as the discrimination loss:
\begin{equation}
\label{equ:dloss}
\dloss=\log D_\Dparam(\inputrating)+\log\big(1-D_\Dparam(\outputrating)\big).
\end{equation}

Inspired by the idea of GAN~\cite{GoodfellowPMXWOCB14}, we aim to unify the
different goals of the generator and the discriminator by letting them play a
\emph{minimax game} via optimizing the objective in Eq.~\ref{equ:minmax}. 
Using the direct generation loss as an example, the overall objective of \ours is presented as follows:
\begin{equation}
\begin{aligned}
\label{equ:minmax}
\min_{\mathbf{\Gparam}}&\max_{\Dparam}\,\, \mathcal{L}=\,\,\mathbb{E}_{u\sim \users} [\log D_\Dparam (\inputrating_u)] \\
&\qquad\qquad\quad-\sum_{u\in\users}\log \frac{\exp\big({S_{\Sparam}(\inputrating^{*})_{u,t}}\big)}{\sum_{j\in \items} \exp{\big(S_{\Sparam}(\inputrating^{*})_{u,j}\big)}}\\
&\qquad\qquad\quad+\mathbb{E}_{\outputrating_v \sim G_\Gparam(\inputrating^{(in)})}[\log\big(1-D_\Dparam(\outputrating_v)\big)]\\
&\qquad s.t., \Sparam=\arg \min \rsloss\big(\inputrating^{*}, S_{\Sparam}(\inputrating^{*})\big),
\end{aligned}
\end{equation}
where $\Gparam$ and $\Dparam$ are model parameters of $G$ and $D$, respectively. $u$ is a real user profile sampled from the user universe, $\outputrating_v$ is a fake user profile generated from the generator distribution $G_\Gparam(\inputrating^{(in)})$.

%% file: 5_learning.tex
\section{Learning}
\label{sec:learning}

Training \ours is not trivial. We will discuss and provide solutions to three critical problems in training \ours in this section.

\subsection{Approximate Discretization}
Firstly, the discretization layer in generator employs a discontinuous step function.
The gradient for step function is not defined at the boundary and is zero everywhere else.
This is problematic in for gradient propagation.

\begin{figure}[t]
\begin{center}
\includegraphics[width=0.9\columnwidth]{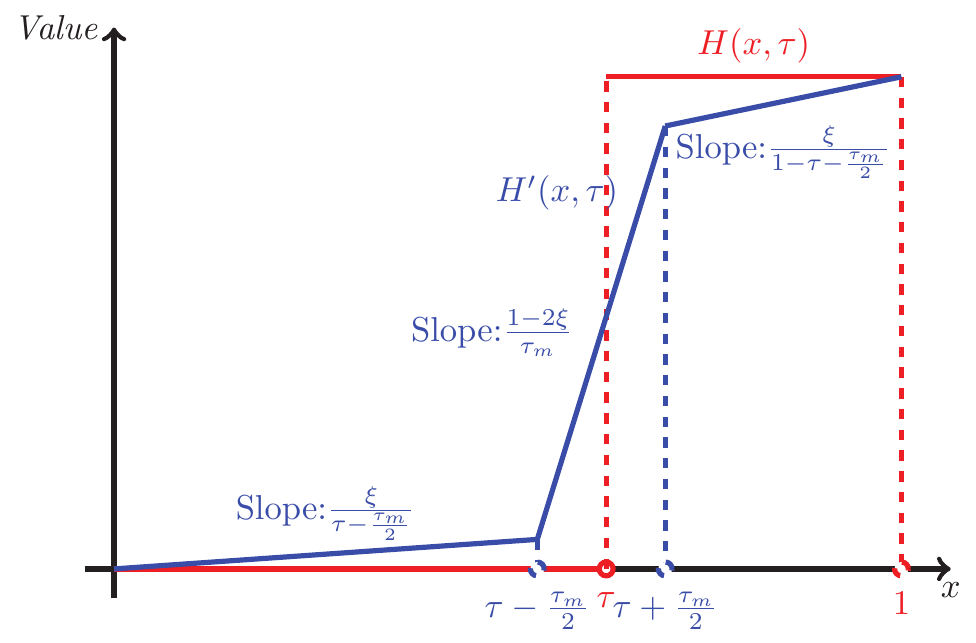}
\caption{We use $H'(x,\tau)$ (line in blue) to approximate the heaviside step function $H(x,\tau)$ (line in red). The approximation contains three straight-line segments with different slope.}
\label{fig:approximate}
\end{center}
\end{figure}

To address this issue, we adopt an approximation function $H'$ to mimic the behavior of the Heaviside step function $H$ introduced in Eq.~\ref{eq:Heaviside}:
\begin{equation}
H'(x,\tau)=
\begin{cases}
\frac{x\xi}{\tau-\frac{\tau_m}{2}}, & x < \tau-\frac{\tau_m}{2}\\
\frac{\xi(x-\tau-\frac{\tau_m}{2})}{1=\tau-\frac{\tau_m}{2}}+1-\tau, & x > \tau+\frac{\tau_m}{2}\\
\frac{(x-\tau)(1-2\xi))}{\tau_m} + 0.5, & others, \\
\end{cases}
\end{equation}
where $\tau_m=\min\{\tau,1-\tau\}$. As shown in Fig.~\ref{fig:approximate}, $H'(x,\tau)$ uses three straight lines to approximate $H(x,\tau)$, when $x\in[0,1]$. As explained by Tsoi et al.~\cite{abs-2009-01367}, a larger value of $\xi$ providers a smoother derivative but further approximation from the Heaviside function. In practice, a value of $\xi=0.1$ works well.

\subsection{Two-level Optimization for $\gloss$}
Secondly, the adversarial loss in Eq.~\ref{equ:gobjective} is computed based on a well-trained surrogate RS model, which requires a two-level optimization procedure.
The exact gradient, with respect to the generator parameters, for the adversarial loss can be written as:
\begin{equation}
\label{equ:gradient}
\partial_{\Gparam} \gloss = \frac{\partial \gloss}{\partial \outputrating}\frac{\partial \outputrating}{\partial \Gparam}+ \frac{\partial \gloss}{\partial \Sparam}\frac{\partial \Sparam}{\partial \outputrating},
\end{equation}
where $\outputrating=G_\Gparam (\inputrating)$.

However, the computation of the exact gradient requires to keep the parameters of the surrogate model for all the training steps, which will consume a lot of time and space resources~\cite{Tang2020Revisiting}.
Following the idea of Tang et al.~\cite{Tang2020Revisiting}, we approximate the gradient by unrolling only the last step when accumulating gradients to improve the efficiency of \ours.

\subsection{Learning Algorithm}
Finally, we present Alg.~\ref{alg:training} to optimize the overall objective of \ours in Eq.~\ref{equ:minmax}.
Specifically, the
generator acts like an ``attacker'' and attempts to generate ``perfect'' fake user profiles that are difficult to detect by the discriminator and can achieve the malicious attack goal on the surrogate model. On the other hand, the discriminator module
performs like a ``defender''. It tries to
accurately distinguish fake user profiles and real
user profiles, and provides guidance to train the generator to generate
undetectable fake user profiles. Each of the generator and the discriminator
strikes to enhance itself to beat the other one at every round of the minimax
competition. 

\begin{algorithm}[t]
\caption{Training procedure for \ours.}
\label{alg:training}
\small
\KwIn{rating matrix $\inputrating$}
\KwOut{parameter set $\Gparam$ for the generator $G$ and parameter set $\Dparam$ for the discriminator $D$}

\For{number of training epochs}
{
    \textbf{(1) Discriminator optimization}
    
	\For{$k_1$ steps}
	{
		uniformly sample a minibatch of users $\mathcal{U}'$\;
		\ForEach{ $u'\in \mathcal{U'}$}
		{
			sample $F$ items to construct $\mathbf{x}^{(in)}_{u'}$\;
		}
		generate a minibatch of fake user profiles $\{\mathbf{x}^{(out)}_{u'}=G(\mathbf{x}^{(in)}_{u'})| u'\in \mathcal{U'}\}$\;
		optimize $\Dparam$ to $\min \dloss$ with $\Gparam$ fixed\;
	}

    \textbf{(2) Generator optimization}
    
	\For{$k_2$ steps}
	{
		uniformly sample a minibatch of user rating vectors $\{\mathbf{x}_{u}\}$\;
		\ForEach{ $u'\in \mathcal{U'}$}
		{
			sample $F$ items to construct $\mathbf{x}^{(in)}_{u'}$\;
		}
		generate a minibatch of fake user profiles $\{\mathbf{x}^{(out)}_{u'}=G(\mathbf{x}^{(in)}_{u'})| u'\in \mathcal{U'}\}$\;
        \textbf{(3) Surrogate optimization}
        
        \For{$T$ steps}
        {
            sample a minibatch of user rating vectors from $(\inputrating,\outputrating)$\;
            optimize $\Sparam$ to $\min \rsloss$;\
        }
        optimize $\Gparam$ to $\min \gloss$ with $\Dparam$ and $\Sparam$ fixed\;
	}
}

\end{algorithm}

%% file: 6_experiment.tex

\section{Experiment}
\label{sec:experiment}

In this section, we conduct experiments\footnote{The source code of \ours is available at \url{https://github.com/XMUDM/ShillingAttack}.} to answer the following research questions:
\begin{itemize}
\item \textbf{RQ1}: Does \ours give better attack performance on different victim RS models, compared with other Shilling Attack methods? (Sec.~\ref{sec:attack})
\item \textbf{RQ2}: Is it more difficult to recognize the attack launched by \ours, compared with other Shilling Attack methods? (Sec.~\ref{sec:detection})
\item \textbf{RQ3}: Does each component in \ours contribute to its attack effects? (Sec.~\ref{sec:ablation})
\item \textbf{RQ4}: Can \ours achieve tailored attack goals such as In-segment Attacks? (Sec.~\ref{sec:secondary})
\item \textbf{RQ5}: How does the attack budget affect the performance of attack? (Sec.~\ref{sec:att_budget})
\item \textbf{RQ6}: Does the choice of the surrogate model affect \ours? (Sec.~\ref{sec:surrogate})
\end{itemize}

\subsection{Experimental Setup}\label{sec:setup}

We use seven benchmark datasets for RS in our experiments, including ML-100K\footnote{\url{https://grouplens.org/datasets/movielens/100k}}, FilmTrust\footnote{\url{https://www.librec.net/datasets/filmtrust.zip}}, Yelp\footnote{\url{https://www.kaggle.com/c/yelp-recruiting/data }} and four other Amazon datasets\footnote{\url{http://jmcauley.ucsd.edu/data/amazon}} Automotive, Tools and Home Improvement (T \& HI), Grocery and Gourmet Food (G \& GF), and Apps for Android (A \& A). 
ML-100K is used as the sole dataset in most previous work on Shilling Attacks~\cite{SandvigMB08} due to the convenience of computation. We use its default training/test split. However, we believe that the nature of datasets affects attack transferability. For example, some victim RS models perform better on denser datasets. Thus, we include the other four datasets, which are larger and sparser, to testify the competence of \ours in different settings. We randomly split datasets except ML-100K by 9:1 for training and testing, respectively.
To exclude cold-start users (as they are too vulnerable), we filter users with less than 15 ratings and items without ratings.
Five target items are randomly sampled for each dataset.
Tab.~\ref{tab:datasets} illustrates the statistics of the data. 
where ``Sparsity'' is the percentage of missing ratings.

\begin{table}[t]
\caption{Statistics of data.}
\centering
\begin{tabular}{|c|c|c|c|c|}
\hline
\textbf{Data} & \textbf{\#Users} & \textbf{\#Items} & \textbf{\#Ratings} & \textbf{Sparsity} \\ \hline
ML-100K       & 943              & 1,682            & 100,000            & 93.70\%           \\ \hline
FilmTrust     & 780              & 721              & 28,799             & 94.88\%           \\ \hline
Yelp          & 2,762            & 10,477           & 119,237            & 99.59\%           \\ \hline
Automotive    & 2,928            & 1,835            & 20,473             & 99.62\%           \\ \hline
T \& HI       & 1,208            & 8,491            & 28,396             & 99.72\%           \\ \hline
G \& GF       & 2,212            & 8,041            & 62,424             & 99.65\%           \\ \hline
A \& A        & 7.845            & 12,615           & 193,928            & 99.80\%           \\ \hline
\end{tabular}
\label{tab:datasets}
\end{table}

\begin{table}[t]
\caption{Attack budget.}
\centering
\begin{tabular}{|c|c|c|c|c|}
\hline
\textbf{Data} & \textbf{$\mathbf{A}$} & \textbf{$\mathbf{P}$} & \textbf{$\mathbf{\left|S\right|}$} & \textbf{\#Targets} \\ \hline
ML-100K       & 50                    & 90                    & 3                                  & 5                  \\ \hline
FilmTrust     & 50                    & 36                    & 3                                  & 5                  \\ \hline
Yelp          & 50                    & 5                    & 1                                 & 5                  \\ \hline
Automotive    & 50                    & 4                     & 1                                  & 5                  \\ \hline
T \& HI       & 50                    & 8                     & 1                                  & 5                  \\ \hline
G \& GF       & 50                    & 10                    & 1                                  & 5                  \\ \hline
A \& A        & 50                    & 9                    & 1                                 & 5                  \\ \hline
\end{tabular}
\label{tab:budget}
\end{table}

The attack budget is shown in Tab.~\ref{tab:budget}. 
We use each attack method to generate $50$ user profiles. This is roughly the population which can manifest the differences among attack models~\cite{BurkeMBW05}. 
In each injected user profile, we demand the number of non-zero ratings to equal to the average number of ratings per user in the dataset. 
This makes the ``profile size'' to be 90, 36, 5, 4, 8, 10 and 9 in ML-100K, FilmTrust, Yelp, Automotive, T \& HI, G \& GF, and A \& A, respectively.

\begin{table*}[ht]
\caption{Analyses of attack performance.}
\centering
	\begin{subtable}{.95\textwidth}
      \centering
        \caption{Use default values for $A$ and $P$.}
        \label{tab:def_ap}
        \begin{tabular}{|c|c|c|c|c|c|c|c|c|}
		\hline
		\textbf{Method}& \textbf{\ours}	&\textbf{AIA}&	\textbf{DCGAN}&	\textbf{WGAN}&	\textbf{Random Attack}&	\textbf{Average Attack}&	\textbf{Segment Attack}&	\textbf{Bandwagon Attack}\\
		\hline
		\#Best results&	30&	2&	0&	0&	0&	1&	8&	1\\
		Ratio (Max: 100\%) &71.43\%&	4.76\%&	0.00\%&	0.00\%	&0.00\%&	2.38\%	&19.05\%	&2.38\%\\
		\hline
		\#Top-2 results&	40&	8&	0&	0&	1&	3&	28&	4\\
		Ratio (Max: 50\%) &	47.62\%&	9.52\%&	0.00\%&	0.00\%&	1.19\%&	3.57\%	&33.33\%&	4.76\%\\
		\hline
		\end{tabular}
		\vspace{5pt}
	\end{subtable}%
    \\
	\begin{subtable}{.95\textwidth}
        \centering
        \caption{Use various values for $A$ and $P$ as in Tab.~\ref{tab:ap}.}
        \label{tab:diff_ap}
        \begin{tabular}{|c|c|c|c|c|c|c|c|c|}
		\hline
		\textbf{Method}& \textbf{\ours}	&\textbf{AIA}&	\textbf{DCGAN}&	\textbf{WGAN}&	\textbf{Random Attack}&	\textbf{Average Attack}&	\textbf{Segment Attack}&	\textbf{Bandwagon Attack}\\
		\hline
		\#Best results&	147&	5&	1&	0&	8&	6&	40&	3\\
		Ratio (Max: 100\%) &70.00\%&	2.38\%&	0.48\%& 0.00\%&	3.81\%	&2.86\%&	19.05\%	&1.43\%	\\
		\hline
		\#Top-2 results&	197&	37&	1&	0&	17&	16&	142&	10\\
		Ratio (Max: 50\%) &	46.91\%&	8.81\%&	0.24\%&	0.00\%&	4.05\%&	3.81\%	&33.81\%&	2.38\%\\
		\hline
		\end{tabular}
	\end{subtable} 
\end{table*}

\begin{figure*}[!t]
\centering
\includegraphics[width=1\textwidth]{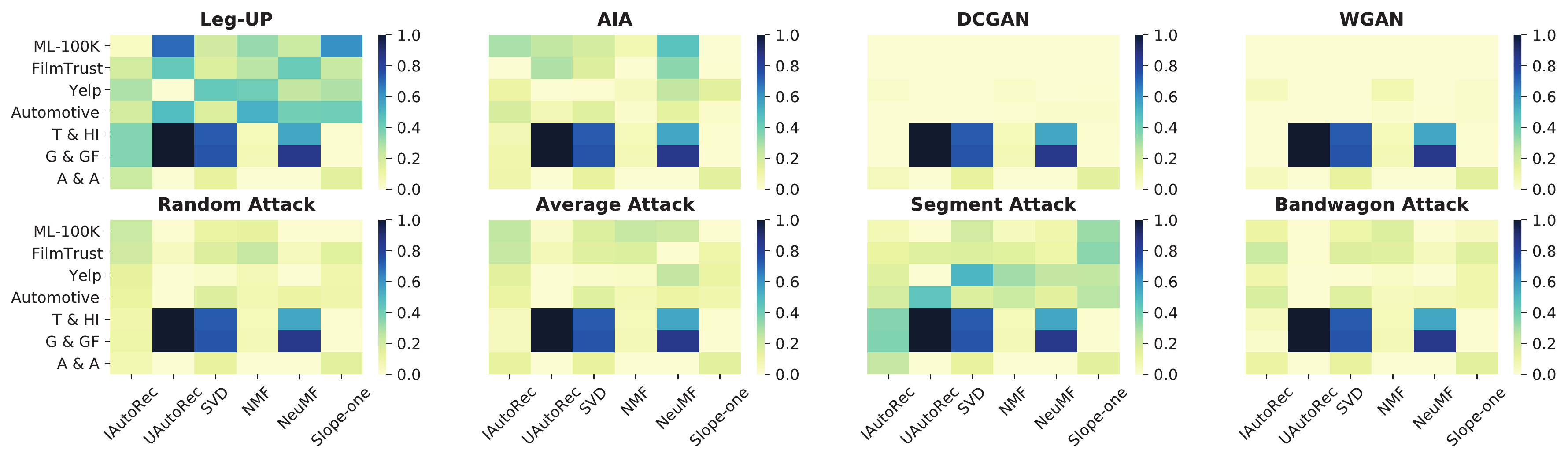}
\caption{Heatmaps showing attack results (default $A$ and $P$). The darker a cell shows, the larger the corresponding $\text{HR}@10$ is.}
\label{fig:attack}
\end{figure*}

\subsection{Attack Effectiveness (RQ1)}\label{sec:attack}
We compare \ours with several Shilling Attack methods, including classic heuristic based methods and recent GAN based methods: 
\begin{enumerate}
\item \textbf{AIA} stands for Adversarial Injection Attack, a bi-level optimization framework to generate fake user profiles by maximizing the attack objective on the surrogate model~\cite{Tang2020Revisiting}. AIA randomly selects $A$ (i.e., the attack size) real user profiles and uses them to initialize the optimization. Following the original paper, we use WRMF as the surrogate model, and unroll the last step 
as an approximation of the adversarial gradient. Note that AIA does not constrain the profile size. It is possible that a generated fake user profile has more than $P$ filler items.
\item \textbf{DCGAN} is a Generative Adversarial Network~\cite{RadfordMC15} adopted in a recent Shilling Attack method~\cite{Christakopoulou19}, where the generator takes a noise and outputs fake user profiles through convolutional units. We follow the settings in~\cite{Christakopoulou19} for the network structures and hyper-parameters.
We randomly sample $P$ non-zero ratings in the fake user profiles for a fair comparison.
\item \textbf{WGAN} is the Wasserstein GAN~\cite{ArjovskyCB17} which has shown better empirical performance than the original GAN in many tasks~\cite{Ishaan2017WGAN}. We replace the GAN architecture in DCGAN with WGAN for Shilling Attacks. The hyper-parameters are the same as  DCGAN.
\item \textbf{Random Attack} is a heuristic based method~\cite{BurkeMBW05}, which assigns random rating $\outputrating_{u,i} \sim\mathcal{N}(\mu,\sigma)$ to $P$ random items in the fake user profile $u$, where $P$ is the profile size, $\mu$ and $\sigma$ are the mean and the variance of all ratings in the system, respectively.
\item \textbf{Average Attack}~\cite{BurkeMBW05} assumes that the fake user assigns a rating $\outputrating_{u,i} \sim\mathcal{N}(\mu_i,\sigma_i)$ to $P$ randomly sampled items, where $\mu_i$ and $\sigma_i$ are the mean and the variance of ratings on this item $i$, respectively.
\item \textbf{Segment Attack} divides the fake user profile into selected items and filler items. It assigns maximal ratings to the selected items and minimal ratings to the filler items. Following~\cite{BurkeMBW05}, for each target item in ML-100K, we select three items that are most frequently rated under the same tag/category of the target item as the selected items. 
For each target item in the other six datasets which do not have information of tag/category, we sample three items (FilmTrust) and one item (Yelp, Automotive, T \& HI, G \& GF, and A \& A) from global popular items as the selected items~\cite{Lin2020Attacking}.
\item \textbf{Bandwagon Attack}~\cite{Burke2005Limited} uses the most popular items as the selected items and assigns the maximal rating to them, while fillers are assigned with ratings in the same manner as Random Attack.
\end{enumerate}
WGAN and DCGAN learn the rating assigned to the target item.
In other attack methods for comparison, the highest possible rating (i.e., 5-star) is assigned to the target item. The hidden layer in the AE (i.e., the generator) of \ours (i.e., Eq.~\ref{equ:AE}) has 128 neurons. The first and the second hidden layers in the discriminator of \ours (i.e., Eq.~\ref{equ:D}) have 512 and 128 neurons, respectively. For AIA and \ours, we use WRMF with $\lambda=10^{-5}$ as the default surrogate model and optimize it with Adam~\cite{KingmaB14}.

We evaluate Shilling Attacks on a wide range of victim RS models before and after the attack, including shallow models (NMF~\cite{Lee2001Algorithms}, Slope-One~\cite{Lemire2005Slope}, and SVD)\footnote{For all shallow models, we use implementations from \url{https://github.com/NicolasHug/Surprise}.} and deep learning based models (NeuMF~\cite{HeLZNHC17} and variants of AutoEncoder~\cite{Sedhain2015AutoRec}, i.e., IAutoRec and UAutoRec)\footnote{For all deep learning based models, we use implementations at \url{https://github.com/cheungdaven/DeepRec}.}.
Note that \ours can generate discrete ratings. Since our goal is to estimate how the rating values in fake user profiles spoof RS models, we modify NeuMF, which is originally designed for implicit feedback, to predict explicit ratings, by using the RMSE loss.

Before attacking, we train each victim RS model on the training set. Then, we
deploy each attack method to generate fake user profiles. The required
information for each attack method (e.g., mean and variance) is obtained from
the training set. After the injection, each victim RS model will be trained
again from scratch on the polluted data. We train victim RS models, attack
competitors and \ours until convergence.

We evaluate the attack performance on the test set using Hit Ratios at $K$
with $K=10$ (i.e., $\text{HR}@10$) on the target item. The larger $\text{HR}@10$ is, the more
effective the attacker is. 

Tab.~\ref{tab:def_ap} illustrates how many times and ratios that different attackers produce the highest $\text{HR}@10$ (maximum possible ratio is 100\%) and the top-2 highest $\text{HR}@10$ (maximum possible ratio is 50\% when an attacker produces top-2 best results for all cases, and the highest $\text{HR}@10$ also counts) on different datasets using default values for $A$ and $P$.
And Fig.~\ref{fig:attack} provides heatmaps showing attack performance of different attackers (using default $A$ and $P$) on different victim RS models on various datasets. In Fig.~\ref{fig:attack}, each cell indicates $\text{HR}@10$ of an attack method attacking a victim RS model on a dataset, and darker colors indicate higher $\text{HR}@10$ values.
In the following, we provide analyses of the attack performance:

\vspace{5pt}
\noindent\textbf{Overall Performance.} 
From Tab.~\ref{tab:def_ap}, we can conclude
that, compared to other attack methods, \ours generally achieves attractive attack performance against all
victim RS models on different datasets. Using default settings of $A$ and $P$, \ours is consistently the best attack method compared to all competitors ($71.43\%$ of the time and the maximum possible ratio is $100\%$), or the top-2 best attack method ($47.62\%$ of the time and the maximum possible ratio is $50\%$).
On the contrary, \emph{conventional attack models do not show a robust attack
performance like \ours, even though they may exceed \ours in a few cases.}
This conclusion can be confirmed by observing Fig.~\ref{fig:attack}: (1) Most cells in the heatmap of \ours are dark (i.e., high $\text{HR}@10$), and they are darker than corresponding cells in other attackers' heatmaps. (2) Baselines' heatmaps have some cells in dark colors, but they all have a large region in a light color (i.e., low $\text{HR}@10$).

\vspace{5pt}
\noindent\textbf{Comparisons between Heuristic Based and GAN Based Methods.}
From Tab.~\ref{tab:def_ap}, we can also find that heuristic based methods are sometimes powerful. For example, Segment Attack is frequently the top-2 best attack method ($33.33\%$ of the time), while \emph{directly adopting the idea of Adversarial Attacks (i.e., using general GANs) does not give satisfactory performance}. Particularly, both DCGAN which is adopted in the recent Shilling Attacks~\cite{Christakopoulou19} and WGAN~\cite{ArjovskyCB17} which aims at stabilizing GAN training do not show better performance than simple heuristic based attack approaches like Average Attack and Random Attack. It shows that we need to tailor the idea of general GAN before applying it in Shilling Attacks.

\begin{figure}[!t]
\centering
\includegraphics[width=1\columnwidth]{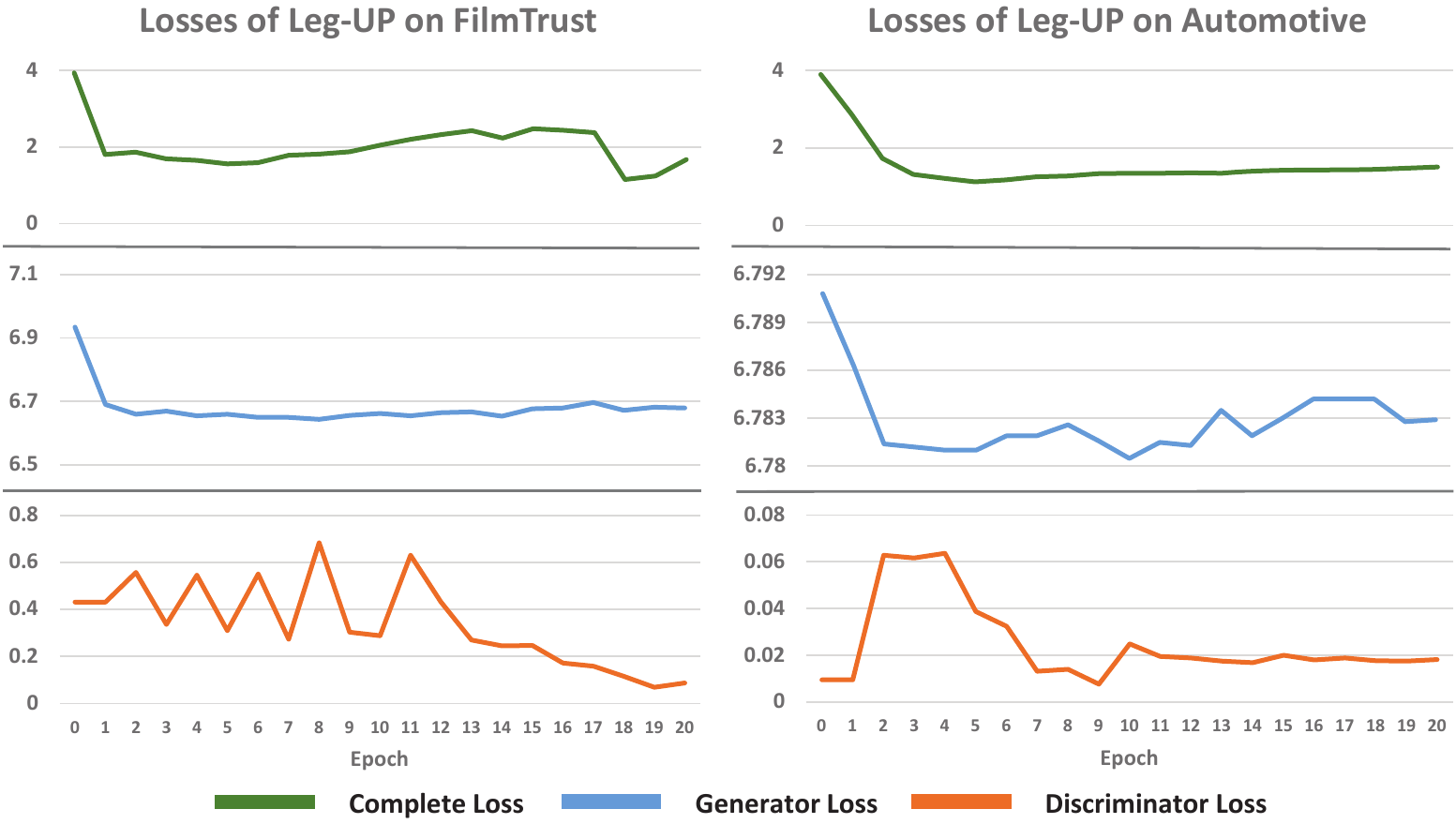}
\caption{Changes of Losses in \ours.}
\label{fig:loss}
\end{figure}

\vspace{5pt}
\noindent\textbf{Analysis of Different Losses in \ours.} 
Fig.~\ref{fig:loss} illustrates how the generator loss, the discriminator loss and the complete loss (i.e., Eq.~\ref{equ:minmax}) of \ours (with default $A$ and $P$) change on datasets FilmTrust and Automotive during the optimization. As the generator and the discriminator compete with each other, we can see fluctuation of their losses, and the increase of one loss is accompanied by the decrease of the other loss. On the whole, we can observe declining trends for both of the generator loss and the discriminator loss, and therefore the complete loss decreases during the optimization.

\subsection{Attack Invisibility (RQ2)}
\label{sec:detection}

\begin{figure}[t]
\begin{center}
\includegraphics[width=0.99\columnwidth]{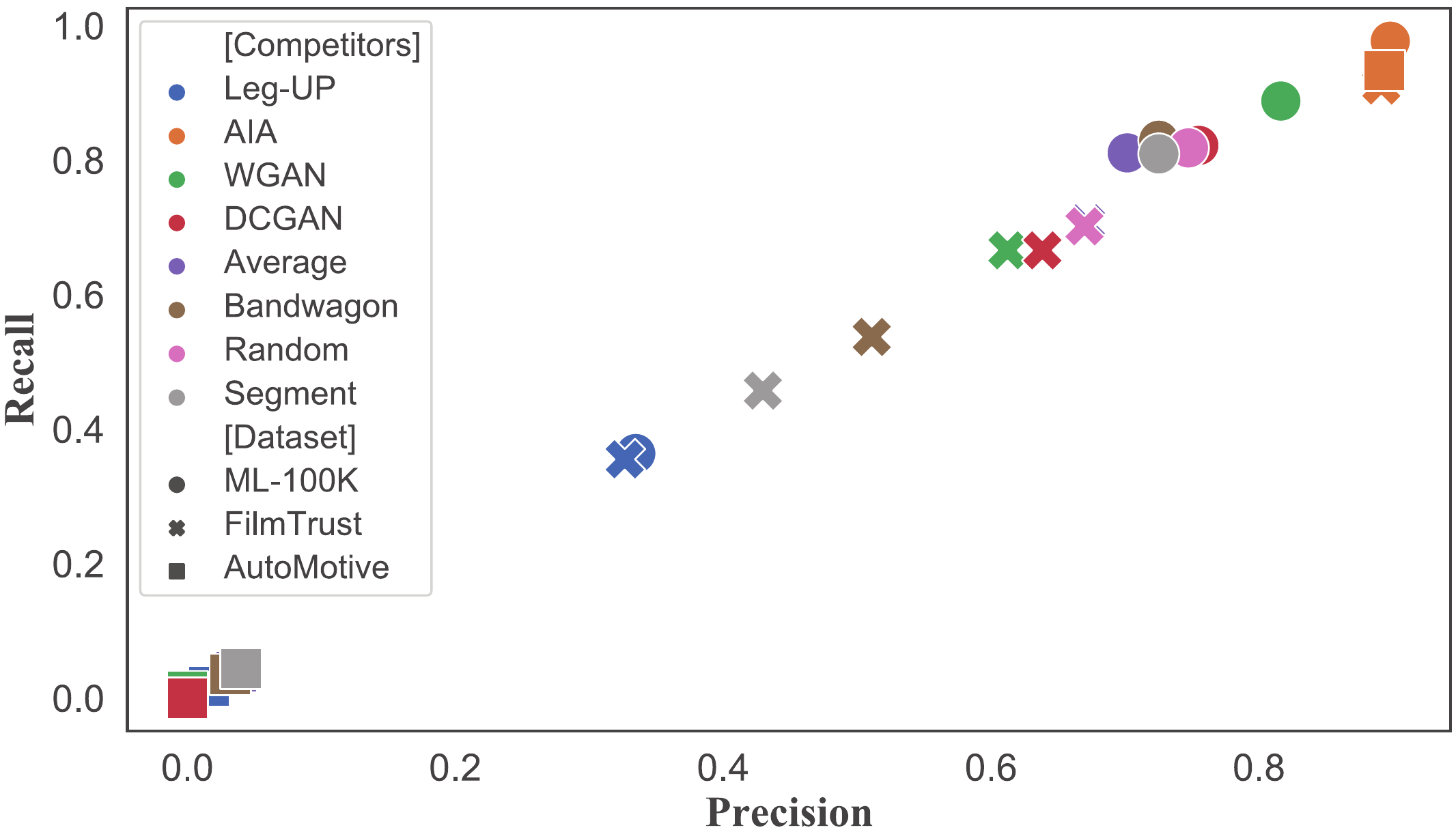}
\caption{Attack detection of injected profiles (precision v.s. recall). Low precision and recall values suggest an invisible attack model.}
\label{fig:detectml}
\end{center}
\end{figure}

\vspace{5pt}
\noindent\textbf{Attack Detection.}
In order to testify how realistic the injected user profiles can be, we apply a state-of-the-art unsupervised attack detector~\cite{Zhang2015Catch} on the injected user profiles generated by different attack methods. Fig.~\ref{fig:detectml} depicts the precision and recall rates of the detector on different attack methods. The lower values of precision and recall indicates that the attack method is more undetectable. We have the following observations based on the detection results: 
\begin{enumerate}
\item There is a positive correlation between precision and recall rates for all attack methods on different datasets.
\item The detection performance is highly dependent on the data and it is easy to detect fake profiles on denser datasets. For example, the detector struggles to detect fake profiles from almost all attackers on Automotive. But it has high precision and recall rates for detecting most attack methods (except \ours) on ML-100K.
\item \ours consistently produces virtually undetectable injections. For most cases, the detector performs the worst against \ours. On the contrary, the detection performance on baseline attackers are unstable. 
\end{enumerate}

\begin{figure}[t]
\begin{center}
\includegraphics[width=0.99\columnwidth]{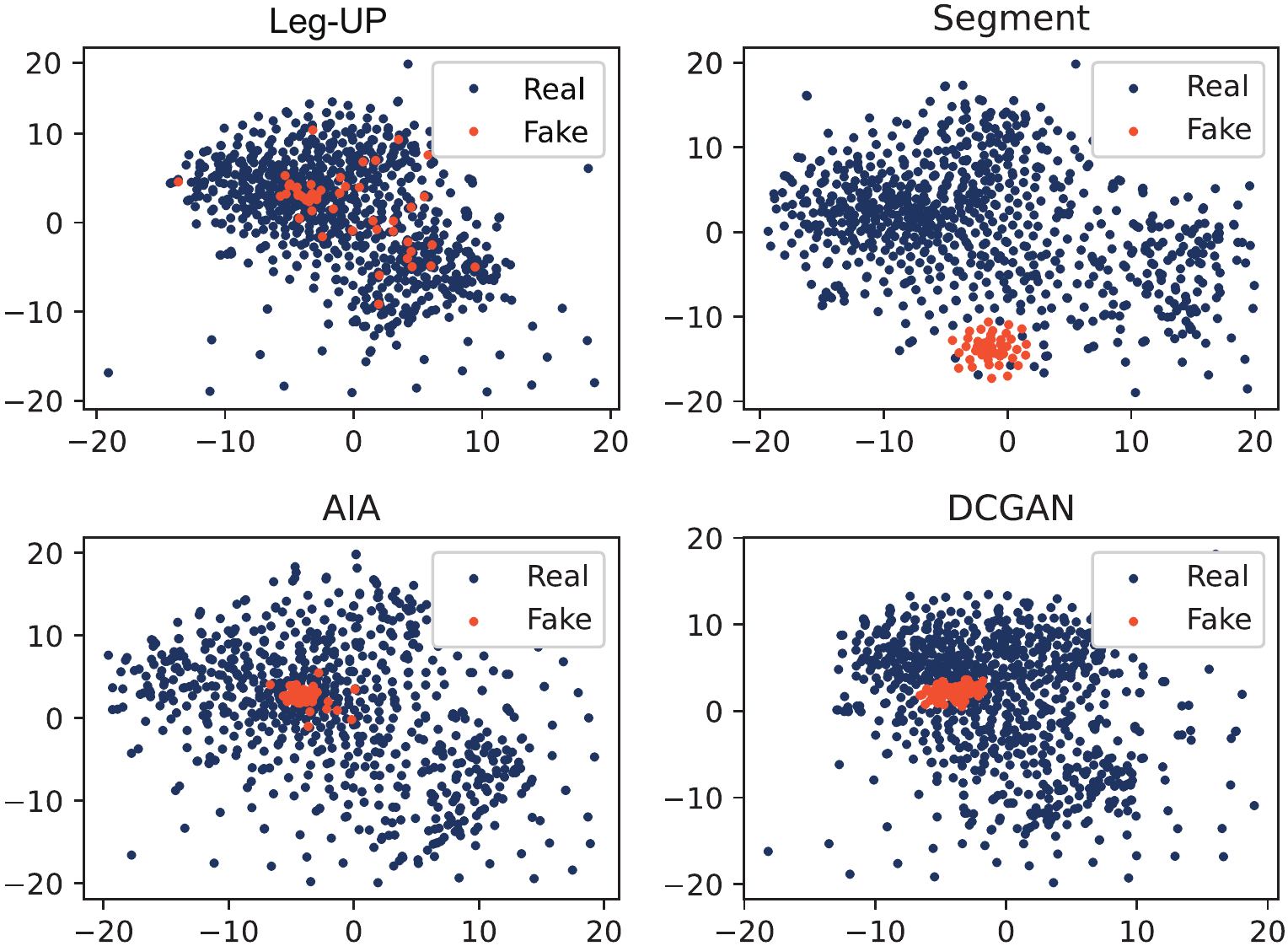}
\caption{Real and fake user profiles in the latent space.}
\label{fig:detectmlvisual}
\end{center}
\end{figure}

\vspace{5pt}
\noindent\textbf{Fake User Distribution.} To further study the attack invisibility of \ours, we visualize the real and fake user profiles using the t-SNE projection~\cite{MaatenH08} of the latent space. Fig.~\ref{fig:detectmlvisual} depicts the visualization results of \ours, Segment Attack, AIA and DCGAN. From the results, we can see that the fake user profiles produced by \ours follow a similar distribution as real user profiles, i.e., fake users are scattered in the latent space of real users. As a comparison, fake users produced by other attackers collapse to a small region in the space, making them easy to detect. Therefore, we can conclude that \ours preserves the diversity of real user behavior patterns in the RS (i.e., personalization), making it more undetectable than other Shilling Attack methods.

\begin{table}[t]
\caption{Ablation study: Impacts of using direct generation loss or indirect generation loss on attack performance and detection performance. Better results are shown in bold.}
\label{tab:abla}
\centering
\resizebox{0.9\columnwidth}{!}{
	\begin{subtable}[t]{0.5\textwidth}
	\caption{Attack performance ($\text{HR}@10$). }\label{tab:abla1}
	\begin{tabular}{lrrrrrr}
	\toprule
	 Dataset         & \multicolumn{2}{c}{ML-100K}       & \multicolumn{2}{c}{FilmTrust}     & \multicolumn{2}{c}{Automotive}    \\
	          & Direct        & Indirect            & Direct       & Indirect            & Direct        & Indirect            \\
	\midrule
	IAutoRec  & 0.0145          & \textbf{0.0990} & 0.7180          & \textbf{0.8333} & \textbf{0.9958} & 0.9955          \\
	UAutoRec  & \textbf{0.6450} & 0.2006          & \textbf{0.9997} & 0.3549          & 0.6369          & \textbf{0.8000} \\
	SVD       & \textbf{0.8575} & 0.5521          & \textbf{0.9588} & 0.9245          & \textbf{0.8776} & 0.8649          \\
	NMF       & \textbf{0.2461} & 0.1849          & \textbf{0.3443} & 0.2821          & \textbf{0.1642} & 0.1466          \\
	NeuMF     & 0.1833          & \textbf{0.3131} & 0.4546          & \textbf{0.6868} & 0.3333          & \textbf{0.3712} \\
	Slope-one & \textbf{0.5835} & 0.0258          & \textbf{0.4061} & 0.0678          & \textbf{0.1133} & 0.0869   \\
	\bottomrule
	\end{tabular}
	\vspace{10pt}
	\end{subtable}
}
\resizebox{0.9\columnwidth}{!}{
	\vspace{10pt}
	\begin{subtable}[t]{0.5\textwidth}
	\caption{Detection results.}\label{tab:abla2}
	\begin{tabular}{crrrrrr}
	\toprule
	Dataset & \multicolumn{2}{c}{ML-100K} & \multicolumn{2}{c}{FilmTrust} & \multicolumn{2}{c}{Automotive} \\
	                           & Direct & Indirect            & Direct  & Indirect             & Direct          & Indirect      \\
	\midrule
	Precision                  & 0.3347    & \textbf{0.1522} & 0.3265     & \textbf{0.3107}  & \textbf{0.0160}    & 0.0600    \\
	Recall                     & 0.3638    & \textbf{0.1830} & 0.3552     & \textbf{0.3238}  & \textbf{0.0176}    & 0.0660   \\
	\bottomrule
	\end{tabular}
	\end{subtable}
}
\end{table}

\begin{table}[t]
\caption{Ablation study: Comparisons between using AutoEncoder (AE) or using a simple function (SF as the generator in \ours. Better results are shown in bold.}
\label{tab:ablb}
\centering
\resizebox{0.9\columnwidth}{!}{
	\begin{subtable}[t]{0.5\textwidth}
	\caption{Attack performance.}\label{tab:ablb1}
	\begin{tabular}{lrrrrrr}
	\toprule
	 Dataset         & \multicolumn{2}{c}{ML-100K}       & \multicolumn{2}{c}{FilmTrust}     & \multicolumn{2}{c}{Automotive}    \\
	          & AE       & SF             & AE        & SF            & AE        & SF             \\
	\midrule
	IAutoRec  & 0.0145          & \textbf{0.0448} & 0.7180          & \textbf{0.8229} & \textbf{0.9955} & 0.9677          \\
	UAutoRec  & \textbf{0.6450} & 0.3644          & \textbf{0.9997} & 0.2561          & 0.6369          & \textbf{0.7886} \\
	SVD       & \textbf{0.8575} & 0.5095          & \textbf{0.9588} & 0.9430          & \textbf{0.8776} & 0.8580          \\
	NMF       & \textbf{0.2461} & 0.1364          & \textbf{0.3443} & 0.1397          & \textbf{0.1642} & 0.1030          \\
	NeuMF     & 0.1833          & \textbf{0.2793} & 0.4546          & \textbf{0.5631} & 0.3333          & \textbf{0.2320} \\
	Slope-one & \textbf{0.5835} & 0.0094          & \textbf{0.4061} & 0.0631          & \textbf{0.1133} & 0.0731\\
	\bottomrule
	\end{tabular}
	\vspace{10pt}
\end{subtable}
}
\resizebox{0.9\columnwidth}{!}{
	\begin{subtable}[t]{0.5\textwidth}
	\caption{Detection results.}\label{tab:ablb2}
	\begin{tabular}{crrrrrr}
	\toprule
	Dataset & \multicolumn{2}{c}{ML-100K} & \multicolumn{2}{c}{FilmTrust} & \multicolumn{2}{c}{Automotive} \\
	                             & AE       & SF             & AE        & SF            & AE        & SF             \\

	\midrule
	Precision                  & 0.3347    & \textbf{0.2974} & 0.3265     & \textbf{0.1845}  & 0.0160    & \textbf{0.0000}    \\
	Recall                     & 0.3638    & \textbf{0.3240} & 0.3552     & \textbf{0.2000}  & 0.0176    & \textbf{0.0000}   \\
	\bottomrule
	\end{tabular}
	\end{subtable}
}
\end{table}

\begin{table}[t]
\caption{Ablation study: Comparisons between using discretization with a learnable layer (DL) or using simple rounding (SR). Better results are shown in bold.}
\label{tab:ablc}
\centering
\resizebox{0.9\columnwidth}{!}{
	\begin{subtable}[t]{0.5\textwidth}
	\caption{Attack performance.}\label{tab:ablc1}
	\begin{tabular}{lrrrrrr}
	\toprule
	 Dataset         & \multicolumn{2}{c}{ML-100K}       & \multicolumn{2}{c}{FilmTrust}     & \multicolumn{2}{c}{Automotive}    \\
	          & DL        & SR            & DL        & SR             & DL        & SR            \\
	\midrule
	IAutoRec  & \textbf{0.0145}  & 0.0049   & \textbf{0.7180} & 0.5379          & \textbf{0.9955}    & 0.8858    \\
	UAutoRec  & \textbf{0.6450}  & 0.4304   & \textbf{0.9997} & 0.9869          & \textbf{0.6369}    & 0.6000    \\
	SVD       & \textbf{0.8575}  & 0.6465   & 0.9588          & \textbf{0.9632} & \textbf{0.8776}    & 0.8392    \\
	NMF       & \textbf{0.2461}  & 0.0082   & 0.3443          & \textbf{0.3471} & \textbf{0.1642}    & 0.0790    \\
	NeuMF     & \textbf{0.1833}  & 0.0162   & \textbf{0.4546} & 0.2450          & \textbf{0.3333}    & 0.2611    \\
	Slope-one & \textbf{0.5835}  & 0.5165   & 0.4061          & \textbf{0.4198} & \textbf{0.1133}    & 0.0441\\
	\bottomrule
	\end{tabular}
	\vspace{10pt}
\end{subtable}
}
\resizebox{0.9\columnwidth}{!}{
	\begin{subtable}[t]{0.5\textwidth}
	\caption{Detection results.}\label{tab:ablc2}
	\begin{tabular}{crrrrrr}
	\toprule
	Dataset & \multicolumn{2}{c}{ML-100K} & \multicolumn{2}{c}{FilmTrust} & \multicolumn{2}{c}{Automotive} \\
	                           & DL        & SR            & DL        & SR             & DL        & SR            \\
	\midrule
	Precision                  & 0.3347    & \textbf{0.1440} & 0.3265     & \textbf{0.1959}  & \textbf{0.0160}    & 0.0360    \\
	Recall                     & 0.3638    & \textbf{0.1596} & 0.3552     & \textbf{0.2130}  & \textbf{0.0176}    & 0.0396   \\
	\bottomrule
	\end{tabular}
		\end{subtable}
}
\end{table}

\begin{table}[t]
\caption{Ablation study: Comparisons between \ours with (w. D) and without (w/o. D) the discriminator. Better results are shown in bold.}
\label{tab:abld}
\centering
\resizebox{0.9\columnwidth}{!}{
	\begin{subtable}[t]{0.5\textwidth}
	\caption{Attack performance.}\label{tab:abld1}
	\begin{tabular}{lrrrrrr}
	\toprule
	 Dataset         & \multicolumn{2}{c}{ML-100K}       & \multicolumn{2}{c}{FilmTrust}     & \multicolumn{2}{c}{Automotive}    \\
	          & w. D       & w/o. D           & w. D       & w/o. D           & w. D       & w/o. D     \\
	\midrule
	IAutoRec  & 0.0145 & \textbf{0.0874}          & \textbf{0.7180} & 0.5663          & \textbf{0.9955}    & 0.9565    \\
	UAutoRec  & 0.6450          & \textbf{0.8364} & 0.9997          & \textbf{1.0000} & \textbf{0.6369}    & 0.0003    \\
	SVD       & 0.8575          & \textbf{0.9323} & \textbf{0.9588} & 0.9582          & \textbf{0.8776}    & 0.8297    \\
	NMF       & 0.2461          & \textbf{0.2771} & 0.3443          & \textbf{0.3580} & \textbf{0.1642}    & 0.0923    \\
	NeuMF     & 0.1833          & \textbf{0.5627} & 0.4546          & \textbf{0.9293} & \textbf{0.3333}    & 0.2807    \\
	Slope-one & 0.5835          & \textbf{0.6292} & \textbf{0.4061} & 0.4059          & \textbf{0.1133}    & 0.0459     \\
	\bottomrule
	\end{tabular}
	\vspace{10pt}
	\end{subtable}
}
\resizebox{0.9\columnwidth}{!}{
	\begin{subtable}[t]{0.5\textwidth}
	\caption{Detection performance.}\label{tab:abld2}
	\begin{tabular}{crrrrrr}
	\toprule
	Dataset & \multicolumn{2}{c}{ML-100K} & \multicolumn{2}{c}{FilmTrust} & \multicolumn{2}{c}{Automotive} \\
	                           & w. D       & w/o. D           & w. D       & w/o. D           & w. D       & w/o. D     \\
	\midrule
	Precision                  & \textbf{0.3347}    & 0.4204 & \textbf{0.3265}     & 0.4250  & \textbf{0.0160}    & 0.0200    \\
	Recall                     & \textbf{0.3638}    & 0.4572 & \textbf{0.3552}     & 0.4528  & \textbf{0.0176}    & 0.0200   \\
	\bottomrule
	\end{tabular}
	\end{subtable}
}
\end{table}

\subsection{Ablation Study (RQ3)}\label{sec:ablation}
To answer RQ3, we remove or change some components of \ours and investigate the performance changes.

\vspace{5pt}
\noindent\textbf{Impacts of Generation Loss.}
As mentioned in Sec.~\ref{sec:implementation}, a direct ``attack-related'' loss (Eq.~\ref{equ:gobjective}) is beneficial to the improvement of the attack performance, while an indirect ``reconstruction-related'' loss (Eq.~\ref{equ:reconstruction}) can improve attack invisibility. 
This is supported by Tab.~\ref{tab:abla}, which reports the attack performance and detection results of \ours (i.e., direct loss), where the generation loss is directly measured upon the attack performance, and AUSH (i.e., indirect loss), where the generation loss is measured upon the reconstruction performance.
We can see that, in most cases, indirect loss is outperformed by direct loss in terms of $\text{HR}@10$.
However, with indirect loss, AUSH can more often generate indistinguishable fake user profiles than \ours.

\vspace{5pt}
\noindent\textbf{Impacts of Generator Architecture.}
\ours can use a simple function or a complex neural network module as the generator. Note that, when using the simple function, \ours resembles AIA, where ratings of ``template'' users are used to initiate the optimization process. When using a neural network module like Autoencoder, the parameters of the neural network instead of the rating values themselves, will be updated during optimization. As shown in Tab.~\ref{tab:ablb}, removing the Autoencoder decreases $\text{HR}@10$. Adopting a non-linear module such as Autoencoder can capture the complex, or even high-order interactions among users and items. As a result, the generated fake user profiles can sabotage the victim RS models and the attack transferability is improved.

\vspace{5pt}
\noindent\textbf{Impacts of Discretization Strategy.}
Since typical RS use Likert-scale ratings, it is necessary to generate fake user profiles with discrete ratings (e.g., one to five stars) to mimic the normal interactions between normal users and RS. \ours provides two strategies: simply using rounding to map numerical values (i.e., user preferences) to discrete stars like AUSH~\cite{Lin2020Attacking}, or learning how to discretize user preferences into discrete ratings. In Tab.~\ref{tab:ablc}, we report results produced by the discretization layer (DL) or the simple rounding (SR). We can see that, discretization with learnable thresholds helps raise $\text{HR}@10$, because it involves the discretization in the optimization of the overall attack framework so that the discretization errors can be reduced during optimization.

\vspace{5pt}
\noindent\textbf{Impacts of Using a Discriminator.}
\ours is built upon the idea of GAN, where the discriminator is incorporated to guide the generator to produce ``perfect'' fake user profiles. To study the impacts of using the discriminator, we report $\text{HR}@10$ by only using the generator in \ours in Tab.~\ref{tab:abld}. We can see that, incorporating the discriminator is usually harmful in terms of $\text{HR}@10$. However, using the  discriminator is indispensable to generate undetectable fake user profiles for \ours. Therefore, we consider it as a flexible option to include discriminator in \ours to launch an effective attack which should be difficult for detectors to discover.

\subsection{In-segment Attack Goals (RQ4).}
\label{sec:secondary}

Like some existing Shilling Attack methods~\cite{Christakopoulou19,Lin2020Attacking}, it is easy to modify \ours to achieve a tailored attack goal, e.g., attacking in-segment users like Segment Attack does.
In-segment users~\cite{LamR04} are users who have shown preferences on selected items. They are popular target audience because companies which face fierce competitions are eager to attract market population of their competitors.
We define in-segment users as users who have assigned high ratings (i.e., $4$- or $5$-stars) on all selected item in the training set. We can modify the generation loss in Eq.~\ref{equ:gloss} so that in-segment attack can be launched by \ours:
\begin{equation}
\gloss=-\sum_{u\in\users^{(seg)}}\log \frac{\exp{\predict_{u,t}}}{\sum_{j\in \items} \exp{\predict_{u,j}}},
\end{equation}
where $\users^{(seg)}$ is the set of in-segment users and $t$ is the target item.

\begin{figure}[!t]
\begin{center}
\includegraphics[width=0.75\columnwidth]{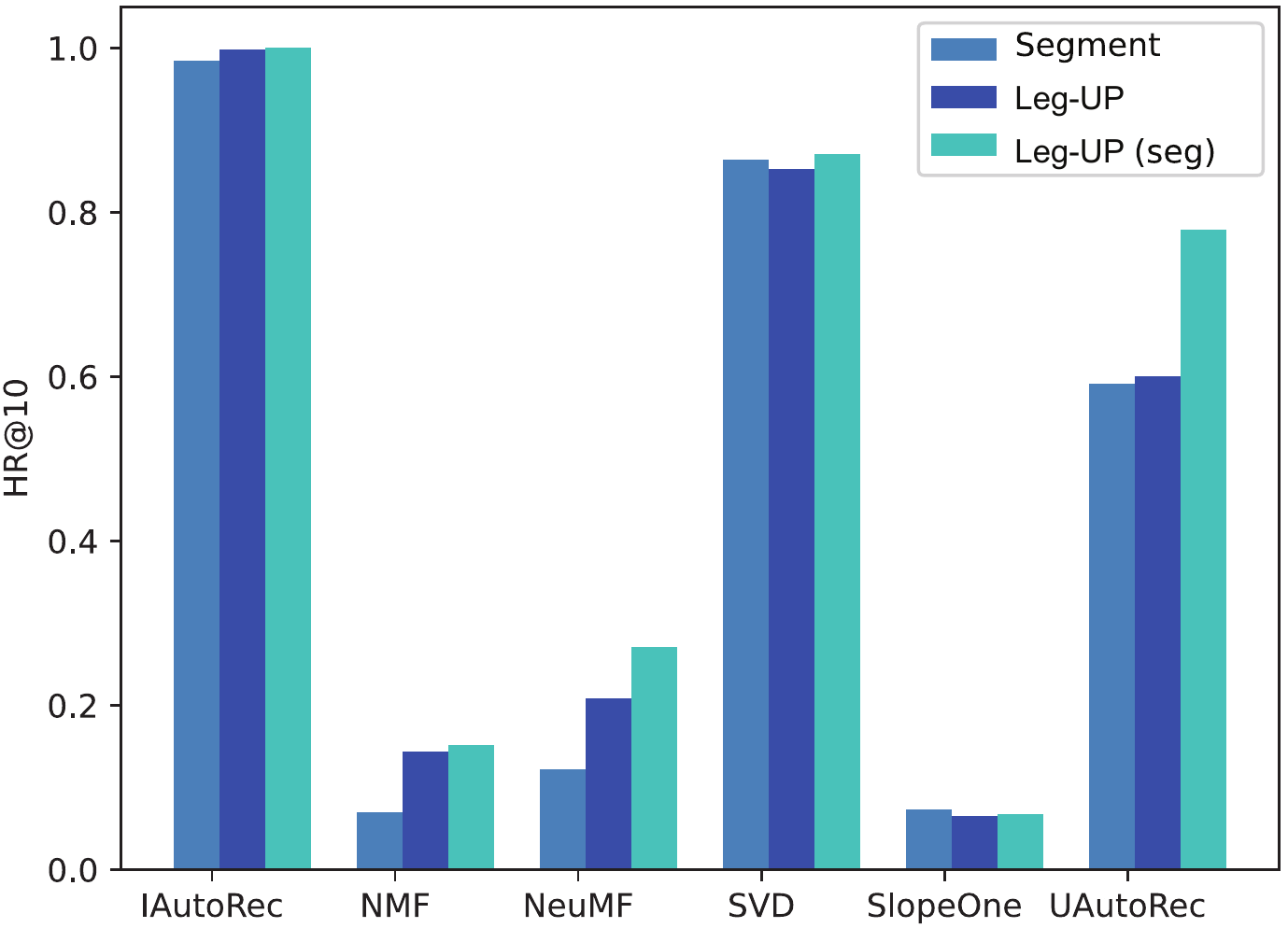}
\caption{$\text{HR}@10$ on in-segment users for Automotive. Higher value suggests a better attack method.}
\label{fig:attackautoseg}
\end{center}
\end{figure}

Fig.~\ref{fig:attackautoseg} reports $\text{HR}@10$ of Segment Attack, \ours and \ours with the tailored loss (i.e., \ours (seg)) against different victim RS models on in-segment users for Automotive dataset. 
We can observe that by modifying the objective, \ours can always obtain higher $\text{HR}@10$ results on in-segment users. We can also find that, \ours generally has better attack performance on in-segment users than Segment Attack.

\subsection{Effects of Attack Budget (RQ5).}
\label{sec:att_budget}

We also investigate the effect of the attack budget hyper-parameter $A$ and $P$ on \ours. In Tab.~\ref{tab:ap}, we report the performance of \ours using varying $A$ and $P$. We can observe that larger values of $A$ improves the attack performance of \ours as more fake user profiles are injected to spoof the system.
For the profile size $P$, \ours achieves reasonable performance using the default values (i.e., the average number of ratings per user in the dataset) on different datasets, justifying the default setting for $p$. Compared to default values of $P$, smaller or larger $P$ values reduce the attack performance of \ours or only slightly improve the attack performance of \ours.

Moreover, we compare \ours to other attack methods using varying $A$ and $P$ (the values are listed in Tab.~\ref{tab:ap}) on different datasets. Tab.~\ref{tab:diff_ap} reports how many times and ratios that \ours produces the highest $\text{HR}@10$ and the top-2 highest $\text{HR}@10$. Similar to the results depicted in Tab.~\ref{tab:ap}, Tab.~\ref{tab:diff_ap} shows that \ours consistently achieves best attack performance for most of the time when different $A$ and $P$ are used in the experiments. Hence, we can conclude that \ours is robust regardless of the setting of the attack budget.

\begin{table}[!t]
\caption{Performance of \ours with varying $A$ and $P$.}\label{tab:ap}
\centering
\scalebox{0.88}{
\begin{tabular}{|c|c|c|c|c|c|c|}
\hline
\multirow{3}{*}{\textbf{ML-100K}}    & \textbf{A}     & 50     & 50     & 50     & 38     & 64     \\ \cline{2-7} 
                                     & \textbf{P}     & 75     & 90     & 110    & 90     & 90     \\ \cline{2-7} 
                                     & \textbf{HR@10} & 0.1550 & 0.1833 & 0.1916 & 0.1501 & 0.2644 \\ \hline\hline
\multirow{3}{*}{\textbf{FilmTrust}}  & \textbf{A}     & 50     & 50     & 50     & 25     & 75     \\ \cline{2-7} 
                                     & \textbf{P}     & 22     & 36     & 50     & 36     & 36     \\ \cline{2-7} 
                                     & \textbf{HR@10} & 0.3639 & 0.4546 & 0.1821 & 0.1035 & 0.6157 \\ \hline\hline
\multirow{3}{*}{\textbf{Automotive}} & \textbf{A}     & 50     & 50     & 50     & 38     & 64     \\ \cline{2-7} 
                                     & \textbf{P}     & 2      & 4      & 8      & 4      & 4      \\ \cline{2-7} 
                                     & \textbf{HR@10} & 0.3224 & 0.3333 & 0.4338 & 0.3847 & 0.3224 \\ \hline\hline
\multirow{3}{*}{\textbf{T \& HI}}    & \textbf{A}     & 50     & 50     & 50     & 38     & 64     \\ \cline{2-7} 
                                     & \textbf{P}     & 5      & 8      & 15     & 8      & 8      \\ \cline{2-7} 
                                     & \textbf{HR@10} & 0.6205 & 0.5488 & 0.6161 & 0.5620 & 0.5230 \\ \hline\hline
\multirow{3}{*}{\textbf{G \& GF}}    & \textbf{A}     & 50     & 50     & 50     & 38     & 64     \\ \cline{2-7} 
                                     & \textbf{P}     & 5      & 10     & 15     & 10     & 10     \\ \cline{2-7} 
                                     & \textbf{HR@10} & 0.8552 & 0.8561 & 0.8522 & 0.8813 & 0.8715 \\ \hline
\end{tabular}
}
\end{table}

\subsection{Effects of Surrogate Models (RQ6).}
\label{sec:surrogate}

By default, we adopt WRMF as the surrogate model in \ours and AIA. We further investigate the effects of  surrogate models by using another two RS models IAutoRec and SVD++~\cite{Koren08}.
Tab.~\ref{tab:surr} reports the attack performance of \ours and AIA on two datasets T \& HI and G \& GF, using default settings for the attack budget $A$ and $P$.
From the results in Tab.~\ref{tab:surr} and Fig.~\ref{fig:attack}, we can observe that different surrogate models affect the attack performance. However, most of the time, changing the surrogate model in one attack method does not affect the performance too much. Moreover, the attack performance of \ours is robust, i.e., it consistently outperforms AIA by a large margin regardless of the adopted surrogate model.

\begin{table*}[!t]
\caption{Performance of \ours and AIA using different surrogate models with default $A$ and $P$.}\label{tab:surr}
\centering
\begin{tabular}{|c|c|c|c|c|c|c|c|c|}
\hline
\textbf{Dataset}         & \textbf{Attacker}       & \textbf{Surrogate Model} & \textbf{IAutoRec} & \textbf{UAutoRec} & \textbf{SVD} & \textbf{NMF} & \textbf{NeuMF} & \textbf{Slope-one} \\ \hline\hline
\multirow{4}{*}{T \& HI} & \multirow{2}{*}{Leg-UP} & IAutoRec                 & 0.7083            & 0.0041            & 0.7167       & 0.0573       & 0.0157         & 0.2301             \\ \cline{3-9} 
                         &                         & SVD++                    & 0.9348            & 1.0000            & 0.7336       & 0.0616       & 0.0169         & 0.4162             \\ \cline{2-9} 
                         & \multirow{2}{*}{AIA}    & IAutoRec                 & 0.1858            & 0.0000            & 0.7052       & 0.0067       & 0.0056         & 0.1481             \\ \cline{3-9} 
                         &                         & SVD++                    & 0.1998            & 0.0000            & 0.7125       & 0.0070       & 0.0062         & 0.1596             \\ \hline\hline
\multirow{4}{*}{G \& GF} & \multirow{2}{*}{Leg-UP} & IAutoRec                 & 0.8664            & 0.7880            & 0.7364       & 0.0615       & 0.0065         & 0.7648             \\ \cline{3-9} 
                         &                         & SVD++                    & 0.8694            & 1.0000            & 0.7401       & 0.0699       & 0.0072         & 0.8264             \\ \cline{2-9} 
                         & \multirow{2}{*}{AIA}    & IAutoRec                 & 0.2009            & 0.0007            & 0.7413       & 0.0035       & 0.0028         & 0.1537             \\ \cline{3-9} 
                         &                         & SVD++                    & 0.2084            & 0.0006            & 0.7400       & 0.0035       & 0.0031         & 0.1474             \\ \hline
\end{tabular}
\end{table*}

%% file: 7_conclude.tex

\section{Conclusion}
\label{sec:con}
Shilling Attack is one of the most subsistent and profitable attack types
against RS. By injecting a small amount of fake user profiles, where each
profile contains ratings on some items, the target items receive more (or
less) recommendations by the victim RS model. We show that two challenges
arise in designing Shilling Attack methods: low attack transferability and low
attack invisibility. To overcome these  challenges, in this paper, we present
a novel Shilling Attack framework \ours based on the idea of GAN. We discuss
and provide different options for the design of the generator module, the
generation loss, the discriminator module and the learning method for \ours.
The experimental results show the superiority of \ours over state-of-the-art
Shilling Attack methods. \ours is effective against a wide range of shallow
and deep RS models on several benchmark RS datasets. Meanwhile, \ours is
virtually undetectable by the modern RS attack detector. \ours, as a new
Shilling Attack method, can benefit the study of secure and robust RS. 

In the future, an interesting research direction is to study the impacts of
surrogate models on attack transferability. We also plan to study the learning
process of \ours for more efficient learning. Furthermore, user feedback in RS
is usually sequential or multi-modal. Therefore, extending \ours to sequential
or multi-modal data is attractive.